\documentclass[a4paper,fleqn]{cas-sc}
\usepackage[english]{babel}

\usepackage{amsmath}
\usepackage{amssymb}
\usepackage{bm} 
\usepackage{graphicx}
\usepackage{algpseudocode}
\usepackage{algorithm}
\usepackage[numbers]{natbib}

\newcommand{\bx}{\bm{x}}
\newcommand{\ty}{\tilde{y}}
\newcommand{\bty}{\bm{\tilde{y}}}
\newcommand{\by}{\bm{y}}
\newcommand{\bz}{\bm{z}}
\newcommand{\bw}{\bm{w}}
\newcommand{\bs}{\bm{s}}
\newcommand{\bmm}{\bm{m}}
\newcommand{\bt}{{\bm{\theta}}}
\newcommand{\bl}{{\bm{\lambda}}}
\newcommand{\btau}{{\bm{\tau}}}
\newcommand{\wt}{{\bm{w}_T}}
\newcommand{\ws}{{\bm{w}_S}}

\definecolor{myblue}{RGB}{12, 12, 158}

\begin{document}
\let\WriteBookmarks\relax
\def\floatpagepagefraction{1}
\def\textpagefraction{.001}

\title[mode = title]{Fundamental problems in Statistical Physics XIV: Lecture on Machine Learning}
\shorttitle{Lecture on Machine Learning}
\author{Aurélien Decelle}[orcid=0000-0002-3017-0858 ]
\shortauthors{A. Decelle}
\address{Departamento de Física, Universidad Complutense, 28040 Madrid, Spain}
\ead{adecelle@ucm.es}
             
\begin{abstract}
The recent progresses in Machine Learning opened the door to actual applications of learning algorithms but also to new research directions both in the field of Machine Learning directly and, at the edges with other disciplines. The case that interests us is the interface with physics, and more specifically Statistical Physics. In this short lecture, I will try to present first a brief introduction to Machine Learning from the angle of neural networks. After explaining quickly some fundamental models and global aspects of the training procedure, I will discuss into more detail two examples illustrate what can be done from the Statistical Physics perspective.
\end{abstract}



\maketitle


\section{Introduction}

The aim of this lecture and these lecture notes is to give a basic introduction to Machine Learning (ML) and to illustrate, by few examples, how Statistical Physics (SP) can bring interesting insights and new methods to understand this field. By ML, I will be more specifically focused on the recent developments of neural networks and their applications to classification and generative tasks. In these notes, I decided to illustrate the work done in Statistical Physics on simple models that do not need a strong background to understand them. In particular, I wanted to avoid as much as possible the use of black-box elements that can often make it harder to understand some fundamental aspects of more complex models.

\section{Why Machine Learning and Statistical Physics ?}

Statistical Physics has traditionally extended its domains of application to other fields beyond physics. It is hard, and maybe not useful, to be exhaustive, so let me focus on a sub-part of Computer Science related topics. Looking back into the past, we can already see many contributions in this field starting in the '80s. At that time, we see the first contributions of statistical physicists to the development and understanding of neural networks, such as the Hopfield model~\cite{amit1985storing} or the perceptron~\cite{gardner1988optimal}. Later on, a lot of work was done in various fields of Computer Science, ranging from constraint satisfaction problems~\cite{mezard2009constraint} to compressed sensing~\cite{donoho2009message,krzakala2012statistical} and more generally Bayesian inference~\cite{nishimori2001statistical}. In fact, in the late '90s, yet another strong connection between the Ising model and Bayesian inference was made. It was shown that the so-called Nishimori line\cite{nishimori1980exact}, a set of coupling constants satisfying a gauge symmetry in a spin glass model, is linked to the Bayesian optimal condition in inference problems\cite{iba1999nishimori}. Furthermore, during the last 40 years, many tools from disordered systems were used, one of the most emblematic being the replica method\footnote{particularly under the light of the Nobel Prize 2021, Giorgio Parisi}, to study and characterize macroscopic properties of various problems, such as the coloring problem~\cite{zdeborova2007phase}, detecting communities~\cite{decelle2011inference}, or more recently, the learning curves of teacher-student neural networks~\cite{loureiro2021capturing} or to characterize the phase diagram of unsupervised learning models~\cite{barra2017phase,tubiana2017emergence,decelle2018thermodynamics}. Therefore, two approaches can be considered concerning the interplay of ML and SP. A first approach would be to consider how ML can be exploited in concrete applications for SP. For instance, many recent works deal with the use of neural networks to classify phase transitions~\cite{carrasquilla2017machine}, or to cluster the phase space automatically. A second approach, the one which interests us, considers how SP can bring new fundamental questions about ML models and new tools to analyze them. Hence, this lecture focuses on showing how methods and questions risen by SP give new perspectives on  ML models. Of course, many previous pedagogical materials already exist following this perspective, and the reader is invited to read other very interesting materials~\cite{nishimori2001statistical,engel2001statistical,opper2001learning,mezard2009information,zdeborova2016statistical,agliari2020machine}

\section{Basic aspects of Machine Learning}

If one would like to be ``caricatural'', we can say that ML consists in defining a family of functions and fitting its parameters to a given dataset. Said in this simple form, it seems to be quite reductive. However, we should agree that this still requires understanding a lot of aspects. First, how does one define what is a good family of functions and why. Second, how do you fit the parameters. And more profoundly: is there a unique set of good parameters ? and if not, how the learning dynamics, the dynamical process adjusting the parameters to find the ``good'' ones, is driven by the dataset? These questions that pop up naturally when first studying the phenomena of adjusting a complicated function to a dataset in order to perform a given task, will, little by little, lead to more profound interrogations such as the geometry of a dataset and of its new (automatically) constructed representation. At some point, we even speak about ``semantic representation'' when trying to analyze how the transformations performed by a ML model can (re)shape a dataset into a set of meaningful categories. In any case, the generic term \emph{latent representation} is typically used to describe the projections that are learned by the models through the learning dynamics. What is exactly a latent representation ? It obviously depends on the considered architecture --- the family of functions used --- and the dataset. We can give some examples. When doing classification at a basic level, a latent representation can be a partitioning of the input space, the one chosen to represent the dataset. Projecting such partitioning into a new space can then be helpful to perform the classification task. When going to more complicated tasks and architectures, such as classifying human faces with deep neural networks, the learned latent features can correspond to a hierarchical representation of the images. First it detects pieces of the faces (eyes, nose, ...) before ``putting'' them together and using all this information to do the classification task. It is sometimes considered that the latent representations are ``feature detector'': for instance, a binary variable turning on when it detects a particular pattern in a given sample. As we will see, they can also be modeled by stochastic variables. \emph{Latent features}, are sometimes associated to the latent representations, to mention the values carried by the weights, which can sometimes be further analyzed. In this lecture, the choice has been taken to first  introduce the field  of ML starting from basic facts of parameters/curve fitting and to briefly explain how to generalize to more complex tasks and models. However, we do not plan to go further than the basic aspects since the objective of these notes is to introduce some simple models of ML where SP has brought new perspectives. Before going in more detail, let's discuss the three classical categories of learning.

\subsection{Learning categories}

Depending on the task and/or the dataset that one considers, three categories of learning methods are usually distinguished:
\begin{enumerate}
    \item \textbf{Supervised Learning:} this formulation indicates that the learning is performed by matching an input $\bx$ together with an output value, $\by$, that can be a set of continuous or discrete values. In this setting, a family of functions is adjusted --- or we say that the model is trained --- such that for every input of the dataset, the function returns the correct output.
    \item \textbf{Unsupervised Learning:} in this formulation, only the set of inputs is given as entry to the model. The typical case is when the likelihood over a set of samples is maximized. Other options can also be designed, such as \textit{self-supervised learning}, where a supervised problem is created out of a generic dataset (with or without label),  for instance by masking part of the data and designing a training where the model should infer the missing values.
    \item \textbf{Reinforcement Learning:} this case is quite different from the other two, and we will not describe it here. It intends to teach an agent to interact with its environment by means of a Markov decision process. Even though recent successes in Artificial Intelligence have shed light on reinforcement learning, see AlphaGo, AlphaStar and AlphaChess, we will not cover this part.
\end{enumerate}

Before entering into the details of supervised and unsupervised learning, I will mention a couple of clear examples for both categories. For the supervised learning, the clearest example is doing a regression. In a regression, you have as input some values $\bx$ that correspond to some varying control parameters, and the outputs $\by$ for which the values are known within some error of measurement. In such a case, one needs to choose first a family of functions, typically linear functions or polynomials of arbitrary degree. Then, the parameters of the family of functions --- e.g. the coefficient of each term of the polynomial --- are adjusted (or learned) using the dataset by typically minimizing the mean square error between the predicted outputs $\hat{\by}$ and the true ones $\by$. Another example, more typical of the ML field, is the task of classification. In that case, the typical family of functions used in ML is given by neural networks (the precise form depends on the designed architecture), and the strong difference is that the output is a category (or an integer).

For unsupervised learning, there are also well-known cases that are generally designed in the Bayesian context. From my point of view, one of the most famous examples is given by the Gaussian mixture model. In this model, the probability density of the dataset is given by a sum of $K$ Gaussian distributions. The parameters of all the Gaussian distributions are then learned by fitting the likelihood of the probability distribution evaluated on the dataset under consideration. The unsupervised nature of this task relies on the fact that no label is needed --- for instance in the case of Gaussian mixture models, the clustering does not depend on the eventually true labels of the clusters---.

\section{Supervised learning}

Let's formalize the problem in the case of supervised learning in order to describe step by step the ingredients of neural networks in that context.  We start by considering a dataset $\boldsymbol{X} \in \mathbb{R}^{N_v \times N_s}$, where $N_v$ is the dimension of the dataset and $N_s$, the number of samples. We then choose a family of functions $f_\bt$, where the dimension of the output depends on the problem considered. For simplicity, let's imagine that we deal with a regression or a binary classification where the output space has only one dimension. In that case, the family $f_\bt$ has a one-dimensional output and the typical support is $\mathbb{R}$ or $[0,1]$ respectively.

\paragraph{Regression:} we can make the hypothesis that the output is given by the following equation
\begin{equation}
  \tilde{y} = \tilde{f}(\bm{x}) + \eta,
  \label{eq:regression}
\end{equation}
where $\tilde{f}$ is the true function relating the input $\bm{x}$ to the true output, $\tilde{y}$, and $\eta$ corresponds to some Gaussian noise with zero mean and variance $\sigma^2$. Since we do not know $\tilde{f}$, we replace it by $f_\bt$ and we will now infer $\bt$ such that eq. \ref{eq:regression} is satisfied. To do that, we should get rid of the noise $\eta$ by averaging it out:
\begin{equation*}
  p(\bx,\tilde{y}|\bt) = \int p(\bx,\tilde{y},\eta|\bt) d\eta = \int p(\bx,\tilde{y}|\eta,\bt) p(\eta) d\eta,
\end{equation*}
with $p(\bx,\tilde{y}|\eta,\bt) = \delta(\eta - (\tilde{y} - f_\bt(\bm{x}))$. We obtain
\begin{equation*}
  p(\bm{x},\tilde{y}| \bt) = \mathbb{E}_\eta\left[\delta(\eta - (\tilde{y} - f_\bt(\bm{x}))\right] = \frac{1}{\sqrt{2 \pi \sigma^2}} \exp\left(-\frac{(\tilde{y}-f_\theta(\bx))}{2 \sigma^2}\right).
\end{equation*}
We can then use the Bayes theorem:
\begin{equation}
  p(\bt|\{\bm{x}^{(m)}\}_m,\{\tilde{y}^{(m)}\}_m) \propto p(\{\bm{x}^{(m)}\}_m,\{\tilde{y}^{(m)}\}_m | \bt) p(\bt)
  \label{eq:regr}
\end{equation}
where we already derived the likelihood given our hypothesis. In the absence of prior $p(\bt)$ on $\bt$, maximizing the left-hand side of eq. \ref{eq:regr}  corresponds to maximizing the likelihood function $p(\{\bm{x}^{(m)}\}_m,\{\tilde{y}^{(m)}\}_m | \bt)$, or equivalently minimizing the inverse of the log-likelihood that is also called the \emph{loss}:
\begin{align*}
  \hat{\bt} &= {\rm argmin}_\bt \left[ \mathcal{L}(\bt,\{\bm{x}^{(m)}\}_m,\{\tilde{y}^{(m)}\}_m) \right] = {\rm argmin}_\bt \left[ -\log \left( p(\{\bm{x}^{(m)}\}_m,\{\tilde{y}^{(m)}\}_m | \bt) \right) \right]  \\
  &= {\rm argmin}_\bt \left[ \sum_m \left( \tilde{y}^{(m)} - f_{\bt}(\bm{x}^{(m)})\right)^2 \right], 
\end{align*}
where $\lVert . \rVert_2$ corresponds to the $\ell_2$ norm. As this expression cannot be minimized easily for any $f_\theta$, a simple way to find $\hat{\bt}$ consists in performing a gradient descent of the loss function $\mathcal{L}$ with respect to the parameters, or weights $\bt$. Writing the gradient descent gives the following learning rules for iterating the values of the parameters $\bt$:
\begin{equation}
  \bt^{(t+1)} = \bt^{(t)} - \gamma \nabla_\bt \mathcal{L}, \nonumber
\end{equation}
where $\gamma$ is the learning rate of the problem. Thus, starting from some initial conditions $\bt^{(0)}$, this dynamics tells us how to update the parameters to reach values that should maximize the likelihood. In practice, in ML the gradient descent is slightly modified to take its ``stochastic'' formulation. This means that the gradient is not taken over the whole dataset, but rather on smaller non-overlapping sub-sets called minibatches. Therefore, each update is performed over the gradient computed on a given minibatch, and when all minibatches have been used to update the weights, it is said that it corresponds to an epoch. This dynamics is called Stochastic Gradient Descent (SGD) when used to minimize a loss. In the context of ML, these are somehow the main ingredients at the heart of the learning dynamics. Even though it might seem quite simple, there are many interesting questions about how the learning depends on the shape of $f_\bt$, the number of parameters or even on the initial conditions of these parameters. We will try to approach some of these questions in section \ref{sec:rbm} in a particular context.

\paragraph{Classification:} for this task, some precautions need to be taken since the labels of the dataset are discrete, as for instance in the case of binary classification $\{0,1\}$. In fact, we see that by using a family of functions where the output is an integer, the derivative will be zero everywhere, and therefore the gradient descent will give us no information to maximize the likelihood (or to minimize the loss). A generic solution is to use a $f_\bt$ defined in $[0,1]$ to describe the probability of belonging to the class 0 (or 1):
\begin{equation}
  p(\bx^{(m)} \in \mathcal{C}_0) = 1-f_\bt(\bx^{(m)}) \;\;\; \text{ and } \;\;\; p(\bx^{(m)} \in \mathcal{C}_1) = f_\bt(\bx^{(m)}). \nonumber
\end{equation} 
From this, we can construct a likelihood function to maximize. For each data $m$, we want to maximize  (w.r.t. the parameters $\bt$) its probability to be classified correctly. Using the true labels $\tilde{y}^{(m)}$ we get
\begin{equation}
	p(\{\bm{x}^{(m)}\}_m,\{\ty^{(m)}\}_m|\bt) = \prod_m f_\bt(\bx^{(m)})^{\ty^{(m)}} (1-f_\bt(\bx^{(m)}))^{1-\ty^{(m)}},  \nonumber
\end{equation}
recalling that the outputs $\ty^{(m)}$ are zero or one. Considering the log-likelihood, we can now define the loss to minimize as
\begin{equation}
  \mathcal{L} = -\left[\sum_m \ty^{(m)} \log(f_\bt(\bx^{(m)})) + (1-\ty^{(m)})\log(1-f_\bt(\bx^{(m)}))\right]. \nonumber
\end{equation}
Again, in such a case, the function cannot be minsimized directly and a gradient descent, or a learning process, is performed to find the parameters that fit best the likelihood.

\subsection{Perceptron: the building block of neural networks}
It is now the time to set a simple example, one of the first neural networks, called the perceptron~\cite{rosenblatt1958perceptron}. This example is important for many reasons. First, this model, despite being simple, is a building block of deep neural networks. Therefore, it is interesting to understand its behavior. Second, its simplicity allows us to understand geometrically how it classifies a dataset. Finally, we can perform two important computations, first we can prove quite easily a useful convergence property. Second, in section \ref{sec:gardner}, we will show how a method from SP allows us to compute the capacity of this model for an artificial dataset. For readers interested into more details, both about the perceptron and its multilayer version, we shall refer to Coolen's book~\cite{coolen2005theory} which is also dealing with the approach coming from SP.

\begin{figure}
    \centering
    \includegraphics[scale=0.3]{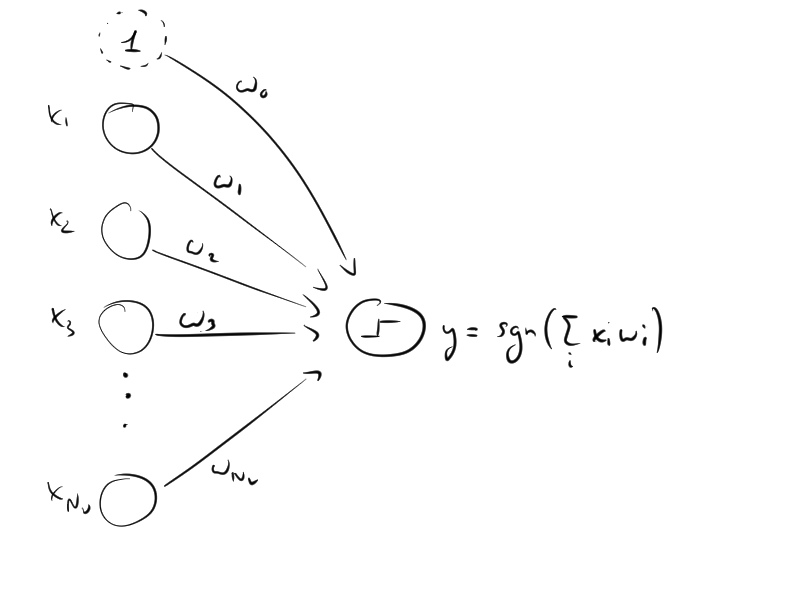}
    \includegraphics[scale=0.4]{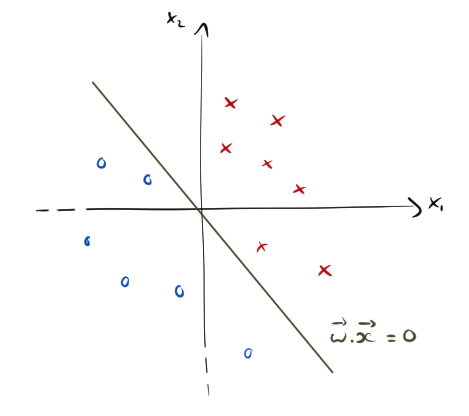}
    \caption{\textbf{Left:} Illustration of the perceptron. The nodes on the left represent the input and the bias. The output is defined by taking the sign of the scalar product of the input and the weights. \textbf{Right:} Illustration of a ``classifiable'' dataset, with the solution separating the two groups of data.}
    \label{fig:perceptron}
\end{figure}

The perceptron is amongst the first models using a neural network architecture. Its design is inspired by the neurons in the brain: a neuron is connected to many other neurons through synaptic connections. It can ``fire'' (be in an excited state) if it received a sufficient number of exciting signals from neurons connected to it, and otherwise it stays at rest. The way it was formalized is as follows. We consider a neuron $y$ connected to a set of $N_v$ neurons $\{x_i\}$. The neuron will be activated with the following rule:
\begin{equation}
  y(\bx)= {\rm sgn}\left(\sum_{i=1}^{N_v} x_i w_i - \alpha\right) \text{ with } {\rm sgn}(x) = \left\{ \begin{array}{c} 1 \text{ if } x \geq 0 \\ -1 \text{ if } x < 0 \end{array} \right., \label{eq:output_perceptron}
\end{equation}
It is common to illustrate this type of model using the picture on the left panel of fig. \ref{fig:perceptron}. We see that in order to activate the neuron $y$, the input signal $\bx$ pondered by the weights $\bw$ should be higher than the threshold $\alpha$, otherwise it remains inactivated. Now, we should understand that a neuron is excited if the input matches a recognized pattern. We can thus classify an input according to the excitation state of the output neuron. For instance, let us imagine that we want to distinguish two classes of objects. We wish to find the parameters $\bw$ such that for a given class the output is always $1$ while for the other class the output is always $-1$. We recognize here the same spirit as for regression tasks: adjusting the weight to obtain a given output. What makes precisely the perceptron a neural network is the choice of the family of functions. In practice, the term neural network is used for families of functions where the output corresponds to a linear operation of the inputs, eventually followed by a non-linearity. In this case, the non-linearity is given by the sign function, but we will see later that more clever choices can be made. 

Before defining the learning rules of the binary perceptron, we can analyze quickly the geometry of the obtained model. We see that, the decision boundary (the set of vectors $\bx$ which are exactly at the border between two classes) corresponds to the solution of the following equation:
\begin{equation}
    \sum_{i=0}^{N_v} x_i w_i  = 0  \Longleftrightarrow \bx \cdot \bw = 0, \nonumber
\end{equation}
where we used the standard notation $x_0=1$ and $w_0 = -\alpha$. We recognize the equation of a hyperplane of dimension $N_v-1$. Hence, the perceptron cuts the $N_v$-dimensional space into two separate parts, where on one side, the output is $1$ and on the other side it is $-1$. From this property, we can already identify a limitation of this model: in order to be able to classify correctly a whole dataset, the representation of the data in the $N_v$ dimensional space should be linearly separable --- there should exist an hyperplane that separates the two classes. See the right panel of fig. \ref{fig:perceptron} for an illustration in two dimensions.

\paragraph{Learning Rule:} in general, the learning rule is derived from the gradient of a loss function. The loss function is often written in terms of some likelihood function or by the classification error. However, in this toy model, the classification is a piece-wise constant function and therefore the gradient is zero almost everywhere. Still, a loss function can be designed for misclassified data points. When dealing with a misclassified data point, we take as loss the distance between the point and the decision boundary (see~\cite{bishop2006pattern,coolen2005theory} for more details). In that case, using the oriented distance, we can write for a given data point $m$:
\begin{equation}
    E(\bx^{(m)}) = \ty^{(m)}\frac{\bx^{(m)} \cdot \bw}{\lVert \bw \rVert}. \nonumber
\end{equation}
Since the denominator is only a multiplicative factor depending on the norm of $\bw$, we can discard it and consider the following gradient
\begin{equation}
    \bm{\nabla}_{\bw} E(\bx^{(m)}) = \ty^{(m)} \bx^{(m)}, \nonumber
\end{equation}
giving the learning dynamics for the data point $(m)$
\begin{equation}
    \bw^{(t+1)} = \bw^{(t)} + \gamma \ty^{(m)} \bx^{(m)}. \nonumber
\end{equation}
where $\gamma$ is the learning rate. Hence, we can write the learning algorithm for the perceptron, see algo. \ref{Algo:perc}.
\begin{algorithm}
    \caption{Learning algorithm for the binary perceptron}
    \label{Algo:GRMM}
    \begin{algorithmic}
        \Statex \textbf{Input:} Data: $\boldsymbol{X} \in \mathbb{R}^{N_v\times N_s}$, Labels: $\boldsymbol{\tilde{Y}} \in \{-1,1\}^{N_s}$, Weights: $\bw^{(0)}$, learning rate $\gamma=1$, $t_{\max}$
        \Statex \textbf{Output:} $\bw^{(*)}$.
        
        \State $\epsilon$ = Nb of misclassified data points.
        \State $T=0$
        \While{$\epsilon \geq 0$ and $t_{\rm max} \geq T$}
        
            \For{$m=0 \rightarrow M$}
                \State Compute $c=\ty^{(m)}(\bx^{(m)} \cdot \bw)$ --- negative for misclassified data
                \If{$c \leq 0$}
                  \State $\bw^{(t+1)} = \bw^{(t)} + \gamma \ty^{(m)} \bx^{(m)}$
                \EndIf
            \EndFor
            \State $T = T + 1$
        \EndWhile
        \State return $\bw^{(*)} = \bw^{(T_{\rm final})}$
    \end{algorithmic}
    \label{Algo:perc}
\end{algorithm}

\noindent Interestingly, it can be proved rigorously that this algorithm converges whenever the data are linearly separable. However, the proof does not tell you how long you need to wait!\footnote{An upper bound is given on the number of updates assuming some hypothesis on the margin that separates the two classes. However, we do not know this margin in general.} This closes the section on the (historical) binary perceptron. In the following, we will see a more classical view of multi-class/multilayer perceptron and where the loss function is defined as a likelihood using a Bayesian approach and where the gradient is naturally given by the derivative of the likelihood of the model.

\subsection{Multilayer perceptron}

Let's first consider the case of a multi-class perceptron using a Bayesian approach, and later we will introduce the possibility of many hidden layers. It is much more convenient to deal with a probabilistic approach to the perceptron. It can be written easily for an arbitrary number of classes, and the loss (to be minimized) can be cast as a maximum likelihood formulation. Also, each data point will be assigned a probability to belong to a cluster instead of a hard assignment. In this formulation, instead of having one output, we have $K$ of them. Therefore, we will call $y_a$ the output, and $\bw_{a}$, the associated set of weights for the neuron $a$. The true label of the dataset will be now a \emph{one hot vector}: if the data $m$ is in class $\ty \in \{0,\dots,K-1\}$, we build the vector $\bty$ which has all components zero but for the one corresponding to its class,  which is put to one. For instance, if the data $m$ is in the class $b$, we will define its label as $\tilde{y}^{(m)}_a = 0$ for $a \neq b$ and $\tilde{y}^{(m)}_b=1$. In particular, we have that $\sum_a \tilde{y}^{(m)}_a = 1$. Now, we define the $K$ outputs of the multilayer perceptron $y_a(\bm{x})$. They are given by a \emph{softmax} activation function:
\begin{equation}
    y_a(\bx) = \frac{\exp(\bx \cdot \bw_a)}{\sum_{c=0}^{K-1} \exp(\bx \cdot \bw_c)} = \frac{\exp(\sum_i x_i w_{ia})}{\sum_{c=0}^{K-1} \exp(\sum_i x_i w_{ic})}.  \nonumber
\end{equation}
We see that the output is normalized, so that it can be interpreted as a probability distribution: they are all positive and sum to one. We can also remark that the ratio of probability between two classes is
\begin{equation}
    \frac{y_a(\bx)}{y_b(\bx)} = \frac{\exp(\bx \cdot \bw_a)}{\exp(\bx \cdot \bw_b)} = \exp(\bx \cdot (\bw_a-\bw_b)) \nonumber
\end{equation}
Again, for a given weight matrix $\bw$, the separation between two classes is given by a hyperplane, this time the equation is $\bx \cdot (\bw_a-\bw_b) = 0$. More precisely, one can demonstrate that any pair of points within a given class, can be linearly interpolated by other points still belonging to the same class.

Once the model is defined, we should decide the loss function that we want to minimize. In this case, we can easily define a likelihood over the model. As for the binary perceptron, we want to maximize the probability that a data is correctly classified,  and maximize it with respect to the model's parameters. The probability that the data $\bx$ is correctly classified according to its label $\tilde{y}$ (or equivalently its one-hot vector $\bm{\tilde{y}}$) is
\begin{equation}
   p(\bx \in \mathcal{C}_{\ty}|\bw) = y_{\ty}(\bx) = \prod_a y_a(\bx)^{\ty_a} . \nonumber
\end{equation}
Taking the log, we obtain the following function to maximize
\begin{equation}
    \mathcal{L} = \sum_m \left[ \sum_a \ty_a^{(m)} \log(y_a(\bx^{(m)}) \right] \nonumber
\end{equation}
This function is also coined as : \emph{binary cross-entropy} (in the case of two classes) and \emph{categorical cross-entropy} in the case of having more than two classes. To maximize the log-likelihood we rely again on the gradient ascent dynamics. The gradient can be easily computed
\begin{equation}
    \frac{\partial \mathcal{L}}{\partial w_{ia}} = x_i \left[ \ty_a - y_a(\bx) \right], \nonumber
\end{equation}
and the learning dynamics hence follow
\begin{equation}
    w_{ia}^{(t+1)} = w_{ia}^{(t)} - \gamma x_i \left[ \ty_a - y_a(\bx) \right]. \nonumber
\end{equation}

\paragraph{Towards multilayer:} we saw that the perceptron with only one input layer connected directly to the output neurons is not capable to classify correctly datasets that are not linearly separable. To overcome this limitation, we will see how the introduction of a hidden layer can solve the problem in a simple example. However, even if we can construct a solution in a simple example, we have no guarantee that the gradient ascent dynamics would converge to it.

\begin{figure}
    \centering
    \includegraphics[scale=0.4]{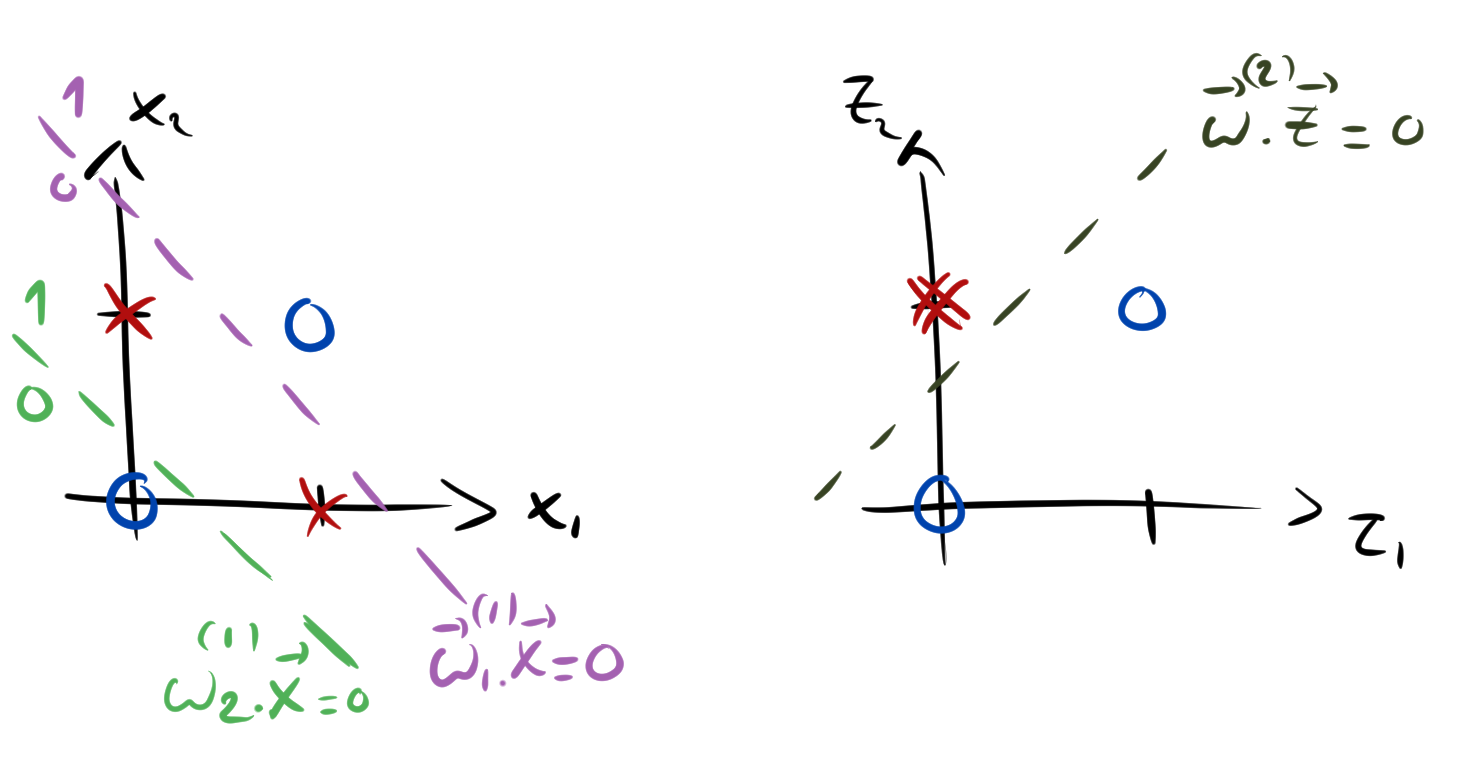}
    \caption{The mechanism of the multilayer perceptron. On the left, we show the XOR dataset in blue and red, which is clearly non-separable by a single hyperplane. In purple and green, the decision surface associated to the two hidden nodes respectively. The latent code of the data, i.e. its representation in the $(z_1,z_2)$ plane, is illustrated on the right panel. In this space, the projection of the dataset is linearly separable, as illustrated by the sketch of one possible decision surface.}
    \label{fig:perceptron_multilayer}
\end{figure}

Let's consider the following dataset: 

\begin{equation}
  \left\{ \begin{array}{l}
  			\bm{x}^{(1)} = \left( \begin{array}{cc} 0 \\ 0 \end{array} \right) \\ \\
  			\tilde{y}^{(1)} = 0
          \end{array} \right., \;\; 
  \left\{ \begin{array}{l}
            \bm{x}^{(2)} = \left( \begin{array}{cc} 1 \\ 1 \end{array} \right) \\ \\
  			\tilde{y}^{(2)} = 0
          \end{array} \right., \;\; 
  \left\{ \begin{array}{l}
            \bm{x}^{(3)} = \left( \begin{array}{cc} 1 \\ 0 \end{array} \right) \\ \\
  			\tilde{y}^{(3)} = 1
          \end{array} \right., \;\; 
  \left\{ \begin{array}{l}
  			\bm{x}^{(4)} = \left( \begin{array}{cc} 0 \\ 1 \end{array} \right) \\ \\
  			\tilde{y}^{(4)} = 1
          \end{array} \right.
\end{equation}
which correspond to the XOR function. It is clear that this dataset cannot be classified correctly by the perceptron since it is not linearly separable, see on fig. \ref{fig:perceptron_multilayer} left panel. We consider now this simple multilayer perceptron. We have two variables $x_1$, $x_2$ as input. Then a set of two hidden neurons $z_1$ and $z_2$ connected to $\bx$ by a set of weights $\bw^{(1)}$ and finally an output neuron $y$ connected to $\bz$ by a set of weights $\bw^{(2)}$. We define now the relation between these variables:
\begin{align*}
  z_1 &= {\rm sig}(\bw_1^{(1)} \cdot \bx), \\
  z_2 &= {\rm sig}(\bw_2^{(1)} \cdot \bx), \\
  y &= {\rm sig}(\bw^{(2)} \cdot \bz).
\end{align*}
Again we considered that the zero component of $\bx$ and $\bz$ is 1 to take into account the biases, and ${\rm sig}(x) = (1+e^{-x})^{-1}$ is the sigmoid function. We can interpret the first layer as the composition of two "simple" perceptrons. In this interpretation, the first hidden layer will cut the input space by two hyperplanes (here lines). Each of the four sub-parts of the plan will then be projected onto a (different) 2-dimensional space with latent code $00$, $01$, $10$ and $11$. The game now is to find the parameters for the two perceptrons $\bw_1^{(1)}$ and $\bw_2^{(1)}$ such that the latent code of the dataset is linearly separable in the latent space of the variable $z$. Then, the last perceptron $\bw^{(2)}$ will be able to classify correctly the dataset. This achieves to explain how a multilayer perceptron can overcome the problem of separability of the dataset, and we can see empirically that the gradient descent does converge in this toy example. See on fig. \ref{fig:perceptron_multilayer} an illustration of this phenomena.

\subsubsection{Multilayer perceptron}

For the sake of completeness, we will derive here the so-called back-propagation algorithm that is used to compute the gradient efficiently. Back-propagation (or chain-rule) is one of the reasons why neural networks can be used in practice, since it allows the computation of the gradient very quickly and to be optimized by linear algebra and parallel operations. Let's deal with the case of one hidden layer, since adding more layers is just as easy but more cumbersome to write. The model of the multilayer perceptron with one hidden layer is designed as follows. We still start from a set of $N_v$ input nodes. Then we add a hidden layer of $N_h$ neurons. Each neuron of the hidden layer will be given by a linear function of the input nodes composed with a non-linearity, also called an \emph{activation function}. We finally have an output layer made of $K$ neurons, each receiving a linear combination of the hidden nodes associated with a softmax activation function. 

\begin{figure}
    \centering
    \includegraphics[scale=0.4]{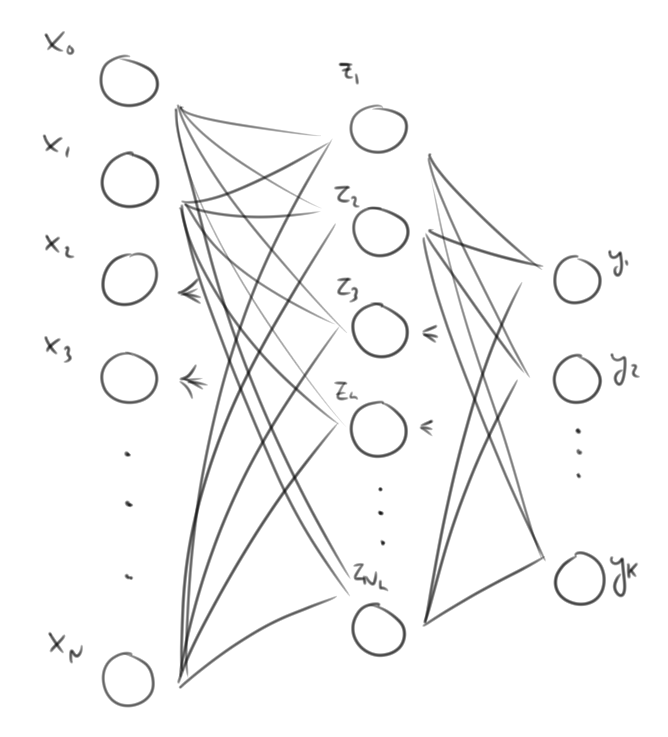}
    \hspace{1cm}
    \includegraphics[scale=0.55]{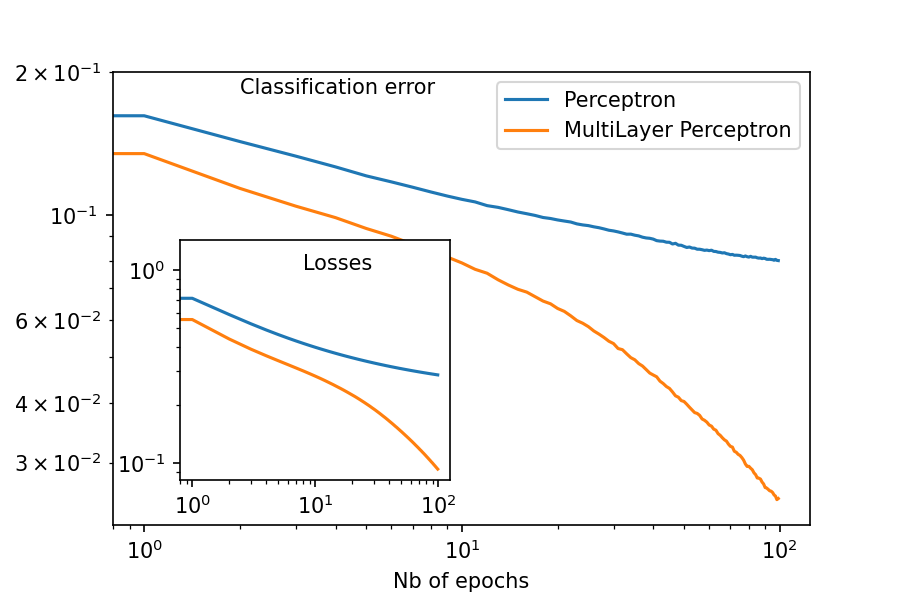}
    \caption{\textbf{Left:} A schematic view of the multilayer (with one hidden layer) perceptron. \textbf{Right:} Learning curve (loss function and classification error) of the perceptron and of the multilayer perceptron with one hidden layer of $N_h =100$ nodes.}
    \label{fig:multilayer}
\end{figure}

Let's note $\bx$ the input vector, $\bz$ the hidden nodes, $\by$ the output layer and $\bw^{(1)}$, $\bw^{(2)}$ the weights between the input layer and the first hidden layer, and the weights between the hidden nodes and the output layer respectively, as can be seen on fig. \ref{fig:multilayer} left panel. The hidden nodes are given by
\begin{align*}
    z_0 &= 1 \\
    z_\alpha(\bx) &= f_1(\bx \cdot \bw_\alpha^{(1)}) \; ; \;\; \alpha=1, \dots, N_h
\end{align*}
where $f_1$ is the activation function. The outputs are given by
\begin{equation*}
    y_a(\bz) = \frac{\exp(\bz\cdot \bw_a^{(2)})}{\sum_{c=1}^K \exp(\bz \cdot \bw_c^{(2)})}.
\end{equation*}
The typical activation functions are the sigmoid or the Rectified Linear Unit (ReLU) function:
\begin{equation}
    {\rm ReLU}(x) = \left\{
    \begin{array}{c} 
        x \text{ if } x \geq 0 \\
        0 \text{ if } x < 0
    \end{array}
    \right. .
\end{equation}
The likelihood of the model is then
\begin{equation}
    \mathcal{L} = \sum_m \left[ \sum_a \ty_a^{(m)} \log\left(y_a(\bz(\bx^{(m)})\right) \right]. \nonumber
\end{equation}
We can now derive the gradient using the chain rule. First, the gradient w.r.t. $\bw^{(2)}$ gives the same expression as for the multi-class perceptron where the inputs are changed for the variables of the hidden layer:
\begin{equation}
    \frac{\partial \mathcal{L}}{\partial w_{\alpha b}^{(2)}} =z_\alpha \left[ \ty_b - y_b(\bz)\right] = z_\alpha \delta^y_b \nonumber
\end{equation}
where we define the variables $\delta^y_b = \ty_b - y_b(\bz)$. A more complicated task is to compute the gradient w.r.t. the weights $\bw^{(1)}$. This can be done without suffering too much using the derivative chain rule.

\begin{align*}
    \frac{\partial \mathcal{L}}{\partial w_{i\alpha}^{(1)}} &= \sum_a \frac{\ty_a}{y_a} \frac{\partial y_a}{\partial w_{i\alpha}^{(1)}}, \\
    \frac{\partial y_a}{\partial w_{i\alpha}^{(1)}} &= \sum_\beta \frac{\partial z_\beta}{\partial w_{i\alpha}^{(1)}} \frac{\partial y_a}{\partial z_\beta} \;\; \text{ with } \frac{\partial y_a}{\partial z_\beta} = y_a w_{\beta a}^{(2)} - y_a \sum_c w_{\beta c}^{(2)} y_c , \\
    \frac{\partial z_\beta}{\partial w_{i\alpha}^{(1)}} &= x_i f_1'(\bx \cdot \bw_\beta^{(1)}) \delta_{\alpha \beta} .
\end{align*}
From this we can obtain the gradient as a matrix-vector product using the variables $\bm{\delta^y}$ defined previously
\begin{equation*}
    \frac{\partial \mathcal{L}}{\partial w_{i\alpha}^{(1)}} = x_i f_1'(\bx \cdot \bw_\alpha^{(1)}) \sum_a w_{a\alpha} \delta^y_a
\end{equation*}
This mechanism remains true for deeper networks: we can write the gradient deep in the network as a matrix-vector product between the previously computed gradient and the corresponding weights at the next layer. Hence, the structure of the gradient enables the possibility to compute it very quickly in computer time (matrix products can be easily parallelized), therefore making these networks trainable in practice. This method has been called \emph{back-propagation}, where it is understood that the error is back-propagated from the output to the rest of the network layer by layer, even though there is nothing more than the chain rule.

We show on fig. \ref{fig:multilayer} right panel an example of training for two cases. In the first one, we take a simple perceptron without hidden layer. In the second case, we take a hidden layer of $N_h=100$ hidden nodes, and again show the behavior of the loss function (inverse of the log-likelihood) and of the classification error: for a data $\bm{x}$, the data is said to be classified correctly if the output with the highest value (or highest probability) of the perceptron corresponds to the correct class. We see that the extra hidden layer is improving substantially the performance, both looking at the accuracy and of the loss.

\subsection{Regression}

\begin{figure}
    \centering
    \includegraphics[scale=0.15]{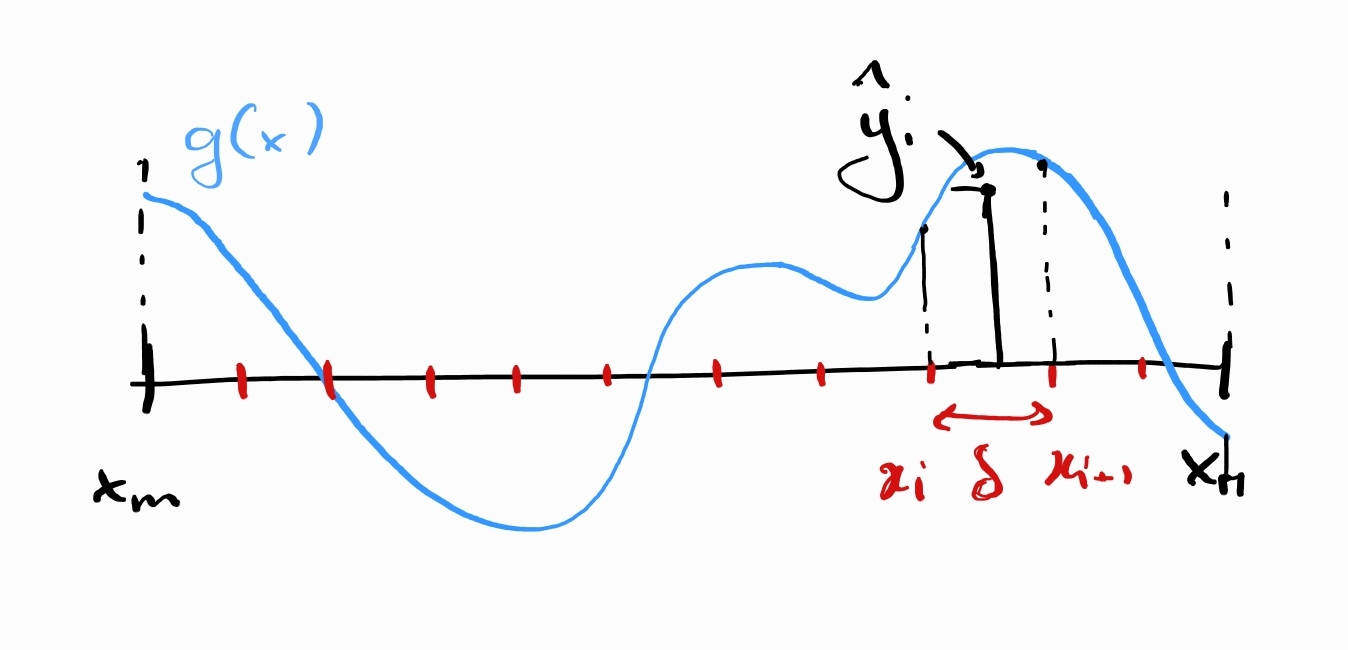}
    \includegraphics[scale=0.5]{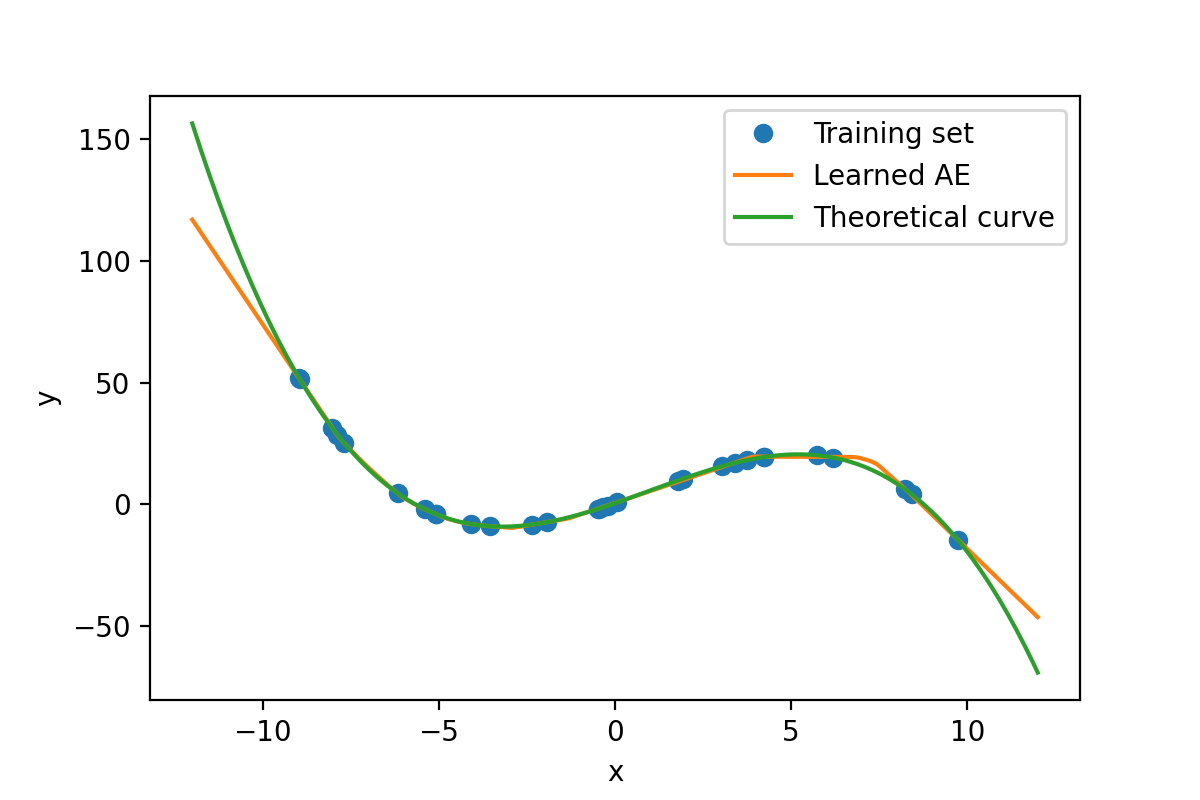}
    \caption{\textbf{Left:} an illustration of the discretization process that is used to encode directly a solution to the neural network. \textbf{Right:} we see the result of the training of a neural network with $N_h=30$ nodes to regress a polynomial. The neural network fits very well the training points, and is quite good at reproducing the trend of the function in this range. As soon as we take values of $x$ outside of the training dataset range, the result diverges quickly with the theoretical function as expected.}
    \label{fig:AE}
\end{figure}

Regression can be dealt with as easily as the classification task. The most notable difference is that the target is usually a continuous value rather than a categorical variable. In this situation, it is enough to adjust the activation function of the last layer of the neural network. A common choice is to use the identity as an activation function such that the output can be mapped into $\mathbb{R}^K$ where $K$ is the dimension of the output. There is no need to enter into more details concerning the computation of the gradient and the learning dynamics since most part is covered already in the previous section. We will take advantage of the regression case to illustrate briefly the fitting properties of neural networks in that case. It can be shown that neural networks with one arbitrarily large hidden layer are ``universal approximators'': that they can fit arbitrarily well any function (as it is the case for the polynomials). To understand this, let me consider a simple example. We imagine that the curve we want to fit is from $\mathbb{R} \rightarrow \mathbb{R}$. Now, imagine that we want to approximate our function $g \in [x_m,x_M]$ by a piecewise constant function. Let's indicate the abscissa at which each continuous piece starts as $x_i$, with $i=1,...,N_h$ and let's take for simplicity $x_{i+1} - x_i = \delta$. For each $x_i$, we will associate the value $\hat{y}_i = (g(x_i) + g(x_{i+1}))/2$, see fig. \ref{fig:AE} left for an illustration. Using this construction, we see that by increasing the number of pieces, the approximation is getting better and better. Now, we can construct explicitly a neural network giving such an output. Consider a neural network with one input layer, one output layer and a hidden layer of $N_h$ hidden nodes, where in this example we will use the Heaviside function for the activation functions. For the first hidden node, we can fix the weight to one and the bias to $-x_m$: $w_{11}^{(1)} = 1$ and $w_{01}^{(1)} = -x_m$. In this way, the neuron will activate as soon as the input passes the value $x_m$. Then, the weight between this node and the output node can be fixed at $w_{1}^{(2)} = \hat{y}_0$ such that it reproduces the first piece of the piecewise continuous approximation. Then, for the second neuron, we repeat the same operation, putting the weight to $w_{12}^{(1)}=1$ and the bias to $w_{02}^{(1)}=-x_m-\delta = -x_1$. Then the weight with the output node will be set to $w_{2}^{(2)} = -\hat{y}_0+\hat{y}_1$ in order to take into account the first activated nodes. It is clear that following this construction, we will reproduce our piecewise continuous approximation defined above with our handcrafted neural network. As stated before, the approximation can be made arbitrarily good by taking higher values of $N_h$ which therefore show that a one-hidden layer neural network can approximate any function, we give a quick example of the results obtained in a simple case explained below and on fig. \ref{fig:AE} right panel. The same is also true for classification.

As an example, we show below the result of a feed-forward neural-network applied to a simple polynomials curve. For this, we use a multilayer neural network associated to a regression task, where the input and output are of dimension one, and that has one hidden layer of dimension $N_h$. The training of the machine is done using a Stochastic Gradient Descent algorithm and the cost function is simply given by the $\ell_2$ norm. We therefore have the following functions for the hidden layer and the output:
\begin{align}
  z_\alpha(x) &= f_1(w_{\alpha 0}^{(1)} + x w_{\alpha 1}^{(1)}) \\
  y(\bm{z}) &= \sum_\alpha z_\alpha w_{\alpha}^{(2)}
\end{align}
where $f_1$ is the activation function for the first layer, here we take $f_1 (x) = {\rm ReLU}(x) = {\rm max}(0,x)$  and the second layer is a linear combination of the hidden entries. The result of our training with $N_h=30$ can be seen on fig. \ref{fig:AE} right panel.

\subsection{Some works on generalization}

Up to now, we only focused on the learning properties of some models according to a set of data on which the parameters are fitted. However, often the ``true'' goal of ML is that the learned model can be applied to other data that were not part of the training set. In other words, to \emph{generalize} its application to unknown data. This new set of unseen data is called \emph{test set}. It is also quite important because we do not wish that our model \emph{badly} overfits the training data, as can occur quite easily when the number of parameters of the model is higher than the number of training samples. While being out of the scope of these lecture notes, it is important to distinguish between overfitting that can be harmful or benign. In the former, the model fits perfectly the training set but gives poor performance on the test set. In the latter, the model is capable of both fitting the training set and performing well on the test set. More on this (maybe) counterintuitive result can be found here~\cite{loureiro2022fluctuations}. In any case, we sketch how the test set is used in order to evaluate a trained model.

In practice, one starts with a dataset $\bm{X}$, having $N_s$ samples. In order to evaluate a learned model and to check for overfitting, the dataset is cut into two pieces. A set of $N_{\rm training}$ samples will be used for training, defining the training set. Then, another set of $N_{\rm test}$ samples, called the test set, is used for measuring performance on unseen samples. These two sets should ideally be disjoint. Usually, something around $80\%$ to $90\%$ of all samples are used for training and the rest for the test. The typical practice is to train the model on the training set, and to evaluate its performance on the test set. In some more complex cases, one has also to fix some meta-parameters of the model, as for instance, the time at which the training is stopped, the value of some regularization parameters, etc. In that case, an additional set, the validation set, is used to fix the optimal value of the meta-parameters and then the model is tested on the test set. This procedure is obviously not perfect but allows defining a clear protocol to compare models.

\subsection{Deep-Learning and Convolutional Neural Networks}

\begin{figure}
    \centering
    \includegraphics[scale=0.2]{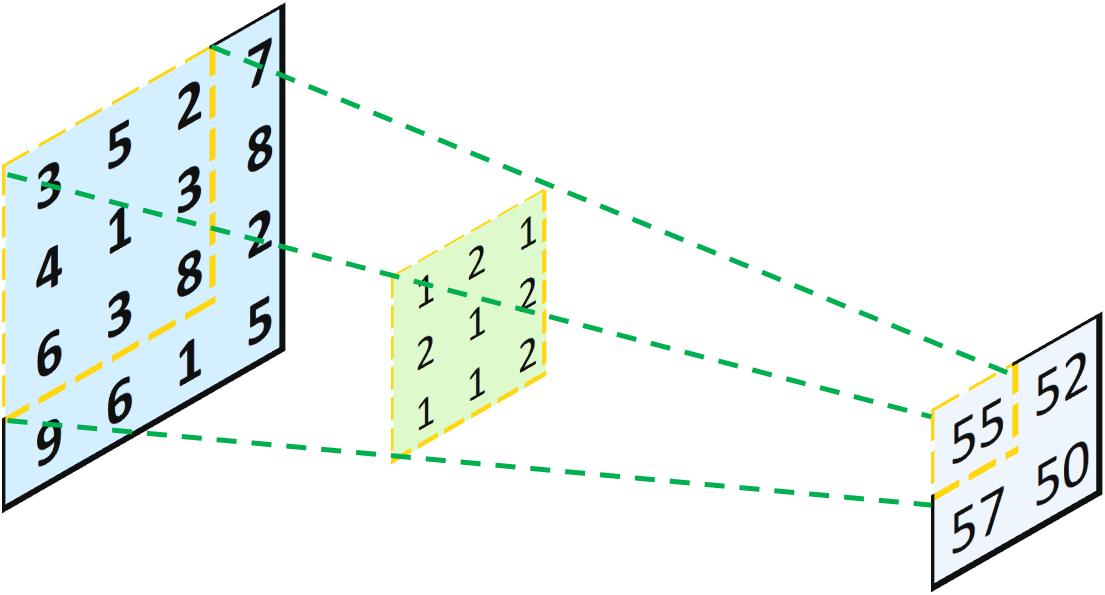}
    \caption{Illustration of a CNN with $3 \times 3$ kernel}
    \label{fig:CNN}
\end{figure}

To conclude with supervised learning, let's briefly describe an important piece of deep-learning: the convolutional neural network or CNN~\cite{lecun1995convolutional}. The CNN was introduced to deal with the translational invariance, particularly for image datasets. In practice, one would like that the features learned by the neural network do not depend on the precise location where they have been detected. A very heavy way to implement this translational invariance would be to add to the dataset every possible translation of it and to train the neural network with this augmented dataset. A simpler way to achieve this is by using CNNs. Let's describe it for the case of images. A CNN consists of a perceptron with a small number of input nodes and that take into consideration the geometry of the input. Let's take the example where the dataset is made of images. We consider a small perceptron having the shape of $5 \times 5$ nodes, thus square shaped. The perceptron will be ``applied'' to the whole image by the following procedure. The square shaped perceptron takes as input a sub-part of $5 \times 5$ pixels of the image and compute the output value. Then, the pixels used as input will be changed, by ``moving'' the perceptron from left to right and up to bottom on the image, spanning it completely and producing an output for each different set of inputs. The important point here, is that all the outputs are made using the same set of weights associated to the CNN, see fig. \ref{fig:CNN} for an illustration. In other words, if we consider the CNN as the following function
\begin{equation*}
    f_{\rm CNN}(x,y) = \left\{
      \begin{array}{cl}
        w_{xy} \text{ if } |x|<=5 \text{ and } |y| <= 5 \\
        0 \text{ otherwise}
      \end{array} \right. ,
\end{equation*}
The number of parameters for this plaquette is therefore $5 \times 5$ to which we add the bias. Then, the output given by the CNN applied to the image is
\begin{equation*}
    z(x,y) = f_1\left( \sum_{x',y'} f_{\rm CNN}(x-x',y-y') {\rm input}(x',y') + b_{xy}\right)
\end{equation*}
where $f_1$ is an activation function and $b_{xy}$ the bias. The CNNs are usually pictured as in fig. \ref{fig:CNN}.
In more general settings, it is also considered that the convolutional kernel can be moved by more than one pixel in each direction, what is called a \emph{stride}. Its default value is often one, but it is possible to shift the kernel using larger values\footnote{Depending on the boundary condition, we can either have an output of the same dimension as the input, or slightly reduced depending on the sliding parameter defined previously, hence reducing the overall dimension.}. At the level of a plaquette, we hope that the neural network will learn a particular filter relevant for the task we are considering. In general, one layer of CNN contains many plaquettes, since we want to detect many different features. Therefore, we will typically have a set of neurons in the CNN-layer scaling as $S_x\times S_y \times N_p$ where $N_p$ is the set of used plaquettes. In practice, the convolutional layer is put together with a 'MaxPooling' layer which is further reducing the output's dimension to prevent overfitting, but it is out of range of these lecture notes to start describing the whole set of possible tricks to train a neural network. In general, it is common to stack many CNN, with possible different shapes to build the architecture. At the end, a typical architecture for Deep-CNN is made of several CNN layers stacked upon the input layer. Before reaching the output layer, few fully-connected layers are added in order to partition the ``features'' detected by the last CNNs.

This concludes my introduction to supervised learning. For a more profound introduction, the reader is invited to look at the literature in Machine Learning, for instance~\cite{bishop2006pattern,goodfellow2016deep}.

\section{Unsupervised learning}

In the unsupervised learning formalism, the setting differs from the supervised case, for the fact that we do not have a target or label variable associated to each data point. In this setting, the general goal is to find a model that reproduces the dataset density distribution. In this section, we will quickly revise the classical models of unsupervised learning based on a neural networks' architecture.

\subsection{Auto-Encoder}
We begin with the AutoEncoder (AE)~\cite{kramer1991nonlinear}, which is a simple way to construct an unsupervised model using neural networks. We will not enter into too much detail of the AE but rather explain its design. As explained before, in the unsupervised learning context, the dataset does not contain any label. Yet, it is possible to construct a neural network without any need of a label. The AE consists in, starting from an input, to perform a set of transformations, starting from the input layer (given by  the dataset) and, as output, a vector of the same dimensions (as the input). We will adjust the parameters of these transformations such that the output is as close as possible to the input. Therefore, given the input $\bm{x}$, we define a function $\bm{y} = \bm{f}_\bt(\bm{x})$ which has the same dimension as the input. In the context of a neural network, the design of $\bm{f}_\bt$ is typically made of hidden layers such as those defined in the supervised section, and convolutional layers. Each hidden layer has its own activation function and for the last layer, the dimension of the output is fixed to be the same as the one of the input while the activation should be adjusted to cover the space in which the dataset is defined. For instance, when dealing with a dataset in $[0,1]$, the last activation function can be the sigmoid, the loss being then the cross-entropy between the input and the output. In the case where the input lives in $\mathbb{R}$, the activation function is the identity and the loss is often the $\ell_2$ norm between the input and the output. As a matter of fact, the AE can be seen as a regressive neural network, where the output nodes are the same ones as the input. With this in mind, one can convince himself that having understood the section on regression is enough to understand the learning mechanism of the AE.

The AE is often seen as a machine able to learn a meaningful and compressed representation of the data~\cite{kramer1991nonlinear,makhzani2015adversarial}. As such, the set of layers chosen in many cases is first operating a reduction of the number of nodes at each layer, and then a set of layers taking the low-dimensional input and bringing it to the same dimension as the dataset's input. The first part is usually referred as the ``encoder'', while the second one is the ``decoder''.

Practical uses of AEs are less clear in comparison to classifiers. To cite few examples, it can be used to automatically find a (hopefully) meaningful representation of the dataset, that can sometimes be used for a classification task later on. It can also be trained to perform a denoising task~\cite{vincent2010stacked}: giving as input a noisy version of a data  but asking the output to be as close as possible of the denoised version of it. It can then perform a denoising task on new data. Finally, it is the basis of the variational-AE that will be briefly discussed in the next section.

\subsubsection{Variational Auto-Encoder}
A few words on Variational Auto-Encoder (VAE)~\cite{kingma2013auto}. VAE is a clever construction to design a latent-variable generative model. A latent-variable model consists in defining a set of latent variables $\bz$, associated to a simple prior distribution $p(\bz)$. And to define the following posterior distribution over the space of the data
\begin{equation}
    p_\bt(\bx) = \int d\bz p_\bt(\bx | \bz) p(\bz)
    \label{eq:like_vae}
\end{equation}
where we need to specify $p_\bt(\bx | \bz)$. A classical approach is to use a neural network to define the conditional distribution as
\begin{equation*}
    p_\bt(\bx | \bz) = \mathcal{N}(\bx|f_\bt(\bz),\sigma)
\end{equation*}
where $\mathcal{N}$ is the Gaussian distribution, and $f_\bt(\bz)$ a neural network taking $\bz$ as input that depends on a set of parameters $\bt$. However, in many cases the posterior distribution $p_\bt(\bx)$ is intractable. Using variational approaches, VAE finds a way to maximize approximately the likelihood eq. \ref{eq:like_vae}. First, let's observe that if the likelihood is intractable, so is the conditional probability $p_\bt(\bz | \bx)$ since it is directly related to $p_\bt(\bx)$ by the Bayes theorem. This would be a problem if we wanted to use expectation-maximization\footnote{for more details see the  \href{https://en.wikipedia.org/wiki/Expectation-maximization_algorithm}{wikipedia page} of EM.} approach as in Gaussian mixture models. Instead, we will introduce a simpler function $q_{\bl}(\bz|\bx)$ to approximate $p_\bt(\bz | \bx)$ which depends on some parameters $\bl$. Now let's write the likelihood of the model
\begin{equation*}
    \log p_\bt (\bx) = \mathbb{E}_q [\log p_\bt] = \mathbb{E}_q \left[\log \left( \frac{p_\bt (\bx,\bz)}{p_\bt (\bz|\bx)} \frac{q_\bl (\bz | \bx)}{q_\bl (\bz | \bx)} \right) \right] = \mathbb{E}_q \left[\log \left( \frac{p_\bt (\bx,\bz)}{q_\bl (\bz | \bx)} \right) \right]  + \mathbb{E}_q \left[\log \left(  \frac{q_\bl (\bz | \bx)}{p_\bt (\bz|\bx)} \right) \right]
\end{equation*}
where $\mathbb{E}_q[.]$ represents the average with respect to the distribution $q_\bl$. The second term corresponds to the Kullback-Leibler (KL) divergence between our ``simple'' distribution $q_\bl$ and the conditional distribution of $\bz$. By definition, it is always positive, and zero if both are equal. Therefore we can rewrite this equation as
\begin{equation}
    \mathbb{E}_q \left[\log \left( \frac{p_\bt (\bx,\bz)}{q_\bl (\bz | \bx)} \right) \right] = \log p_\bt (\bx) - \mathcal{D}_{KL}[q_\bl (\bz | \bx) || p_\bt (\bz|\bx)] \label{eq:vae}
\end{equation}
And we see that, by maximizing the left-hand side of eq. \ref{eq:vae} with respect to all the parameters of the problem: $(\bt,\bl)$, we are at the same time increasing the likelihood of our distribution and decreasing the KL divergence between the true posterior $p_\bt(\bz|\bx)$ and $q_\bl$.
At this point, the term ``variational'' of the VAE is clear, but we still need to explain why it is an AutoEncoder. The reason is that, one can interpret $q_\bl(\bz|\bx)$ as the encoder part of the AE, transforming an input $\bx$ into its latent representation $\bz$. While the decoding part is represented by $p_\bt(\bx|\bz)$, transforming a latent representation $\bz$ into the data-space. More information on practical details and the implementation can be found here~\cite{kingma2013auto,doersch2016tutorial,kingma2019introduction}.

\subsection{Restricted Boltzmann Machine} \label{ssec:rbm}
The Restricted Boltzmann Machine (RBM)~\cite{Smolensky,hinton2002training} is a very appealing model for statistical physicists as it can be seen as a bipartite Ising (or spin glass) model. Its architecture is very similar to a simple AE, however the training procedure is very different.
First, the RBM is a probabilistic model defined on the space where the dataset lives, and a set of hidden nodes. The usual goal when dealing with an RBM is to learn its parameters such that the equilibrium distribution described by our model is a good approximation of the empirical data distribution, If successful we hope that we will be able to generate new data statistically similar as the ones of the dataset (but obviously different from it). The training procedure is based on the maximization of the likelihood of our probabilistic model evaluated on the dataset. 

\begin{figure}[ht!]
    \centerline{\resizebox*{0.7\textwidth}{!}{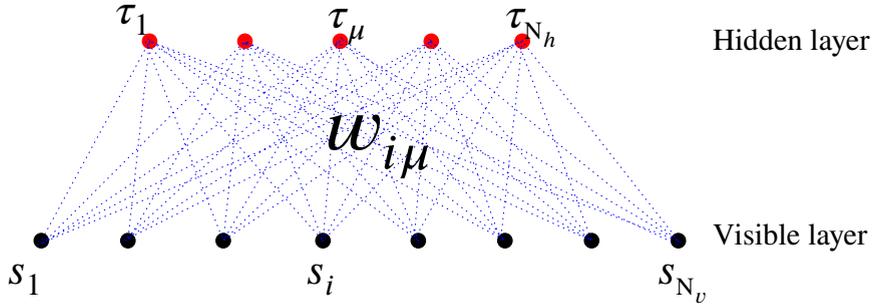}}
    \caption{\label{fig:RBM} Bipartite structure of the RBM.}
\end{figure}

Let's define the model. First, we need to define the input space of the RBM: the visible nodes: $s_i$ with $i=1,\dots,N_v$, and a set of hidden nodes $\tau_\mu$, with $\mu = 1,\dots,N_h$. Then, we define the following Hamiltonian
\begin{equation}
  \mathcal{H}[\bs,\btau] = -\sum_{i,\mu} s_i w_{i\mu} \tau_\mu - \sum_i s_i \theta_i - \sum_\mu  \tau_\mu \eta_\mu ,
\end{equation}
where, in the classical setting, the visible and the hidden nodes are binary discrete variables $\{0,1\}$. The weight matrix $\bw$ represents the couplings between the visible and the hidden nodes, while the parameters $\bt$ and $\bm{\eta}$ are local biases (or local fields for physicists). In practice, any prior distribution can be taken for the visible and the hidden nodes. For instance, the choice of binary $\{\pm 1\}$ is completely equivalent to the previous choice, but it is also possible to consider categorical variables, or even continuous ones using the Gaussian distribution. We note that, if taking a Gaussian prior for the hidden nodes, the distribution over the visible nodes can be mapped to a Hopfield model, see appendix \ref{sec:equivRBM_HOPF}. More discussions on how the prior distribution over the variables shapes the model can be found here~\cite{decelle2021restricted}. This Hamiltonian describes a bipartite Ising model where interactions are present only between the visible and the hidden nodes, but are absent within each layer, see fig. \ref{fig:RBM}. 

The idea behind the RBM is that the correlations between the visible nodes, which ideally would match those of the dataset after learning the weights of the model, will be captured by the interactions between the visible and the hidden layer. This is somehow a different approach from the one of the Boltzmann Machine where interactions between visible nodes are directly present in the model and should modelize the true network architecture. The marginal distribution over the visible nodes is

\begin{equation}
 p(\bs) = \sum_{\btau} p(\bs,\btau) =  \frac{1}{Z} \exp\left( \sum_i s_i \theta_i + \sum_\mu \log[1+\exp(\sum_i w_{i\mu} s_i + \eta_\mu)] \right)
\end{equation}
It is easy to see that this distribution include (potentially) interactions between the visible nodes at all possible orders, for instance by making an expansion of the log. This means that a RBM having discrete variables for the hidden layer can describe a system with higher (than-two) order interactions without the need of specifying exactly which ones, as opposed to the Boltzmann Machine. 

The popularity of this model, both for the field of ML and for physicists, came for its ability to learn the empirical density of complex datasets. In fact, it can be shown that when both the visible and hidden nodes are binary, it is possible to find a set of weights reproducing any discrete distribution (defined on a set of binary variables)~\cite{freund1991unsupervised}. The learning procedure based on the maximum likelihood approach is described into more details in section \ref{sec:rbm}.

\subsection{Generative Adversarial Networks}
The Generative Adversarial Network\cite{goodfellow2014generative} (or GAN) is a generative model based on a competitive training between two separate neural networks. In particular, GANs take advantage of the whole deep-learning machinery. A first neural-network is built called the Generator. The Generator's goal is to generate data similar to the ones of the dataset. They take as input a random number $\bm{x}$ from a simple distribution $p(\bx)$ (from which it is easy to sample). The input is then transformed by a family of functions $g_\bt$ (which is usually a neural network even though it is not mandatory). The Generator maps the input space toward the output space $\{\by\}$ following a set of non-linear transformations. We can define the probability distribution of the Generator as follows
\begin{equation}
    p_G(\by) = \int d\bx p(\bx) \delta(\by - \bm{g}_\bt(\bx)).
\end{equation}
In the general case, this probability distribution is intractable and thus cannot be maximized using ML methods. The training procedure for such a model is therefore particular. To train the Generator to transform the random variables of the prior distribution to data similar to those of the dataset, we use an adversarial approach. A second neural network is built named Discriminator. The Discriminator's task will be to distinguish between dataset samples and data  generated  by the Generator. To do so, it takes as input either a sample from the dataset or a generated one, and, as output, gives a probability of belonging to the true dataset. In general, a different neural network  $d_\bl$ is used for the Discriminator. The training procedure works in a two-steps procedure. In a first step, the Generator will adjust its weights such that it is able to fool the Discriminator. In a second step, the Discriminator will adjust its weights to distinguish better between the true and the generated samples. The following loss is used for the training of the Generator:
\begin{align*}
    \mathcal{L}_G = \int d\by d\bx p(\bx) \delta(\by - \bm{g}_\bt(\bx)) \log\left[1-d_\bl(\bm{g}_\bt(\by)\right] = \int d\by p_G(\by)\log\left[1-d_\bl(\by)\right]
\end{align*}
After many steps of minimizing the loss of the generator (there are many recipes for this, but we will not enter in detail here) the loss of the Discriminator is minimized. The Discriminator's task is to get better and better at distinguishing the true samples from the generated ones. To this purpose, it needs to minimize the following loss
\begin{align*}
    \mathcal{L}_D = -\int p_{\rm dataset}(\bm y) \log\left[ d_\bl(\by) \right] - \int d\by p_G(\by)\log\left[1-d_\bl(\by)\right].
\end{align*}
In this second step, only the weights of the Discriminator are adjusted, while the Generator is fixed. We also see that by the alternance of losses, and by changing in each case only the weights of one of the two neural networks, we can use the loss of the Discriminator and change the sign when optimizing for the Generator.

The whole optimization scheme is quite unstable and, in fact, it is in general not easy to design a working GAN for your specific dataset. Therefore, if possible, it is wise to first find a working GAN in the literature and start to adjust the architecture from it.

This ends the ML introduction, with the definition of some well-known models. We switch now to the SP part, where we will tackle these problems from a different perspective.

\section{Perceptron: Computing the capacity}
\label{sec:gardner}

We will show in this section how to compute the capacity in the perceptron under some hypothesis over the dataset following Statistical Physics arguments --- or equivalently analyzing the typical behavior of the machine --- rather than by looking at the convergence properties of the learning dynamics. In this context, the model was first analyzed by Gardner in the '80s~\cite{gardner1988optimal,gardner1988space}. We make the same kind of computation following the development of~\cite{engel2001statistical}, but also refer to the book of Coolen et al.~\cite{coolen2005theory}. Let me summarize what are the important questions on which we will focus on, and the formalism that we will use to answer them:
\begin{itemize}
    \item Can we compute the generalization error for a perceptron trained with a given number of training samples ?
    \item Can we compute the capacity --- the number of patterns that can be retrieved/learned ?
\end{itemize}
To assess these questions, we shall do a ``typical case'' analysis. That is, at the contrary of the worst case analysis that intends to understand the properties of the most difficult case, we will analyze the typical case, the one that we would face typically assuming some statistical ensemble. The analysis is quite different from the ML approach. We will not at all focus on the learning dynamics, which is still of course an important question. But we will rather look at a Teacher-Student (TS) scenario where we will define an a priori distribution for the considered dataset and for the teacher, and try to look at the typical realizations and analyze their properties. 

\subsection{Teacher-Student analysis}
In the TS scenario, we first define a \emph{teacher} perceptron, chosen from a distribution of possible teachers. The teacher will be an oracle providing us with the correct answer when classifying the training set. We will call the parameters of the teacher $\bm{w}_T$, and note the classification it provides as $y_T(\bm{x}) = {\rm sgn}(\bm{x} \cdot \bm{w}_T)$. The main objective here will be to analyze the statistical properties of the \emph{student}, another perceptron which intends to classify correctly the training dataset provided by the teacher. The parameters of the student will be its weights $\bm{w}_S$. The output of the student is then given by $y_S(\bm{x}) = {\rm sgn}(\bm{x} \cdot \bm{w}_S)$.

Let's add the following notation, the number of inputs will be denoted as $N_v$, and the chosen training set will follow the distribution 
\[
 p(\bm{x}) = \prod_i \left[ \frac{1}{2}\delta_{x_i,-1} + \frac{1}{2}\delta_{x_i,1} \right] .
\]
In general we will consider a training set of $P$ samples together with the classification given by the teacher:
\begin{align*}
    \{\bm{x}^{(m)},\ty^{(m)}\}_{m=1,\dots,P} \,\,\, \text{ where } \,\,\, \ty^{(m)}={\rm sgn}(\bm{x}^{(m)} \cdot \bm{w}_T).
\end{align*}
We can now define the error committed by the student on the training set, and the generalization error as

\begin{align*}
    \epsilon_t(\bm{w}_S,\{\bm{x}^{(m)}\},\bm{w}_T) &= \frac{1}{P}\sum_m \theta(-y_T(\bm{x}^{(m)})y_S(\bm{x}^{(m)})), \\
    \epsilon_G(\bm{w}_S,\bm{w}_T) &= \sum_{\{\bm{x}\}} p(\bm{x})   \theta(-y_T(\bm{x})y_S(\bm{x})).
\end{align*}
where $\theta(x)$ is the Heaviside function. We see that the training error corresponds to the ratio of misclassified samples evaluated from the training set, while the generalization error is computed with respect to the true distribution of the data. 

\begin{figure}
    \centering
    \includegraphics[scale=0.55]{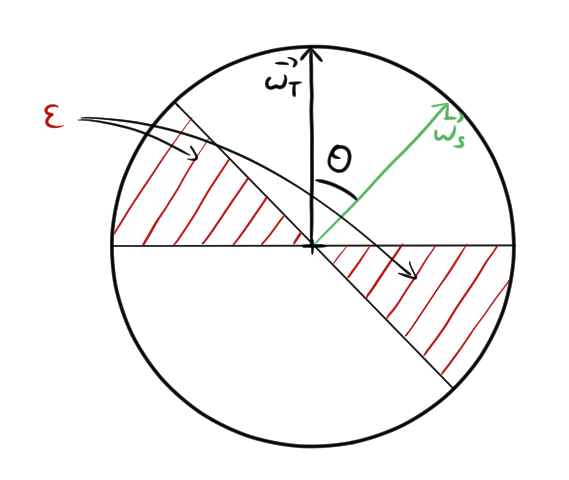}
    \includegraphics[scale=0.55]{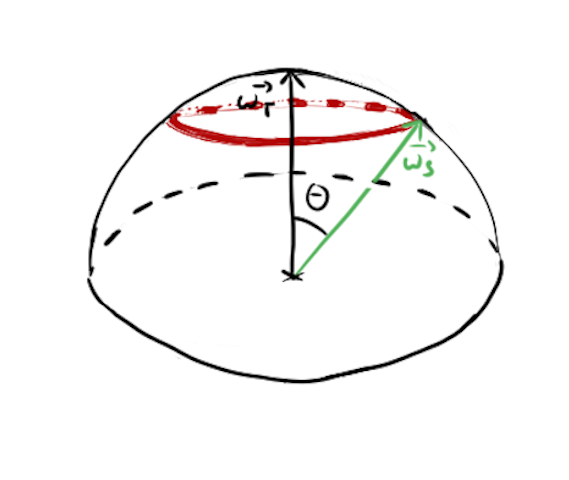}
    \caption{\textbf{Left:} graphical representation of the generalization error given a student vector $\bw_S$ and a teacher $\bw_T$. \textbf{Right:} visualization of the volume of possible perceptrons with a given generalization error in red.}
    \label{fig:sphere_misclass}
\end{figure}

Our typical case analysis will rely on what is called ``Gibbs learning''. That is: instead of learning the weights of the perceptron using a stochastic gradient descent, we look at the space defined by all the possible students that classify correctly the training dataset and take one uniformly at random. Then, we will compute the generalization error of these ``typical students''. To simplify the computation, we will consider that the parameters of the perceptron have a spherical constraint: $\lVert \bm{w}\lVert^2_2 = N_v$: they will live on the hyper-sphere in $N_v$-dimensions. This is only a minor constraint since the classification does not depend on the norm of the weights. With this construction, all data points whose projection form an angle of less than $\pi/2$ with the teacher are classified $+1$, and the others $-1$. Therefore, if the student has an angle $\theta$ with the teacher, we can represent geometrically the volume of misclassified data as on fig. \ref{fig:sphere_misclass} left panel.
Now, the generalization error given the angle $\theta$ is given by the ratio of all possible data points over the ones ending in the misclassified regions. The generalization error is thus given by $\epsilon_G= \theta / \pi$ where $\theta$ is the angle between $\wt$ and $\ws$. We can express the generalization error in terms of the scalar product of the two perceptrons, writing $R=\wt \cdot \ws / N_v$ it gives $\epsilon_G =  {\rm arccos}(R)/\pi$.

\paragraph{The generalization error ---} We wish to answer the question: by taking uniformly at random a student perceptron classifying correctly the training set (amongst all the possible ones), what is the value of the generalization error ? To do that, we need to compute the volume of all the perceptrons that classify correctly the training set. For that, we will make a large deviation computation. To begin, we can compute the volume of all student perceptrons sharing an overlap $R$ with the teacher (hence fixing the generalization error), as illustrated on fig. \ref{fig:sphere_misclass} right panel. Doing this, we find
\begin{equation}
    \Omega_0 (\epsilon) = \int d\ws \delta(\ws^2 - N_v) \delta(R-\cos(\pi \epsilon)) \sim \exp\left(\frac{N_v}{2}[ 1+\log(2\pi)+\log(\sin^2(\pi \epsilon))]\right), \label{eq:perceptron_cap}
\end{equation}
where the details of this computation can be found in appendix \ref{sec:hypersphere}. We see that this entropy will be exponentially dominated by the saddle point $\epsilon^* = 1/2$. It means that, the number of perceptrons giving a classification error of $1/2$ is so huge, comparing to those having a smaller error, that it is impossible to find by chance a perceptron doing better than that. Therefore, taking a student uniformly at random will inevitably gives a result equivalent to randomly guessing the sign of $y$.

\begin{figure}
    \centering
    \includegraphics[scale=0.5]{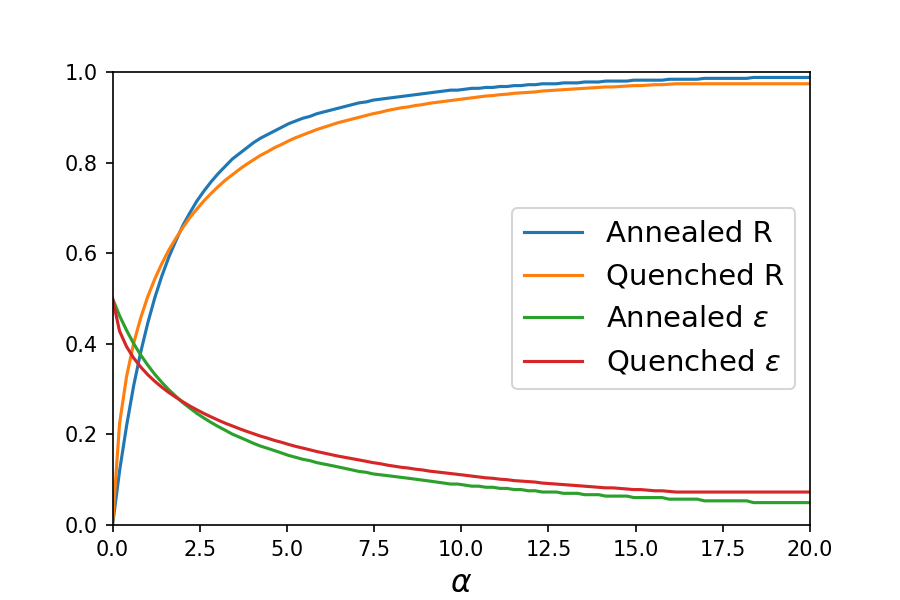}
    \includegraphics[scale=0.5]{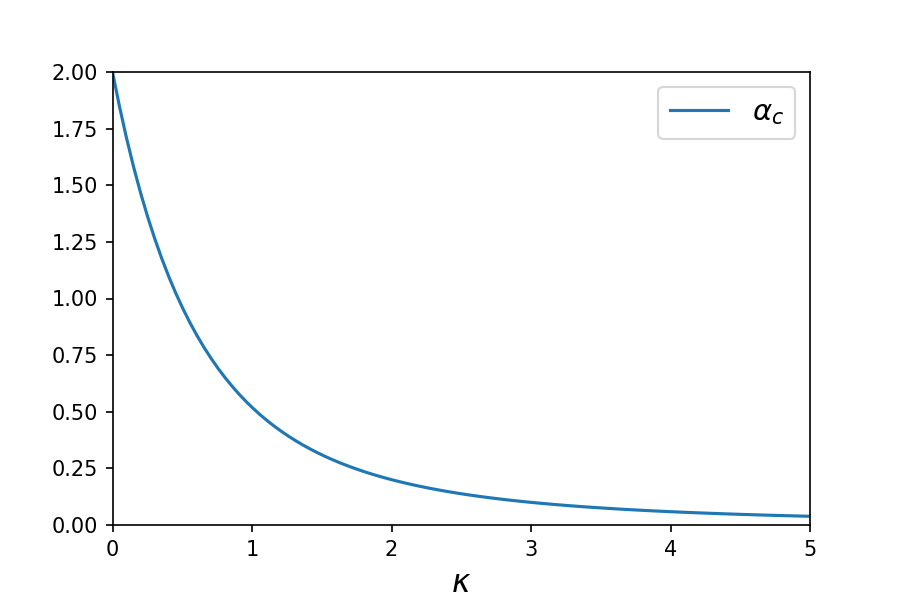}
    \caption{\textbf{Left:} The overlap $R$ and the generalization error in the perceptron obtained as a function of $\alpha=P/N_v$, in both the annealed and quenched cases. \textbf{Right:} The critical capacity of the perceptron as a function of the margin $\kappa$.}
    \label{fig:R_Err}
\end{figure}

Now we can try to see what would be the effect of adding training samples. We add the constraint that the student parameters should classify correctly a number $P$ of samples drawn randomly from the sample's a priori distribution. In our case, we will consider that the samples are uniformly distributed. The quantity is expressed as
\begin{equation*}
    \Omega(\wt,\{\bx^{(m)}\}) = \int \frac{ d\ws \delta(\ws^2 - N_v)}{\left(\int d\ws' \delta(\ws'^2 - N_v)\right)} \prod_m \theta\left[\left(\frac{1}{\sqrt{N_v}} \wt \cdot \bx^{(m)}\right) \left(\frac{1}{\sqrt{N_v}} \ws \cdot \bx^{(m)} \right) \right].
\end{equation*}
where the denominator is here to normalize the whole expression. This expression depends on a particular realization of the training set and of the teacher parameters. In order to obtain the true quenched entropy, we need to compute
\begin{equation*}
    S = \langle \log(\Omega(\wt,\{\bx\}^{(m)})) \rangle_{\wt,\bx \sim p(\bx)},
\end{equation*}
where the average is performed over the training set and all possible teachers. This term is in general hard to deal with. A first simpler approximation can be computed by considering the annealed case
\begin{equation*}
    S_P^{\rm ann} = \log\left( \langle \Omega(\wt,\{\bx\}^{(m)}) \rangle_{\wt,\bx \sim p(\bx)} \right) = \log(\Omega_P)
\end{equation*}
where $P$ indicates the number of considered training samples. The annealed case can be seen as an approximation of the true quenched computation. It is also true that in some cases, when the typical values of $\Omega(\wt,\{\bx\}^{(m)}))$ are very peaked, the annealed computation gives the correct result. However, this usually happens in the high temperature regime (our analysis here is made at zero temperature) --- we will compare later the difference between the annealed and the quenched derivation. The details of the computation can be found in the appendix \ref{sec:comp_entropy} and we give the main steps here for the annealed computation:
\begin{enumerate}
    \item We introduce the Lagrange multipliers $\lambda_m = \frac{1}{\sqrt{N_v}} \ws \cdot \bx^{(m)}$ and $\tau_m = \frac{1}{\sqrt{N_v}} \wt \cdot \bx^{(m)}$ by means of a set of delta functions.
    \item We use the Fourier representation of the delta functions that have been added, introducing the conjugate parameters $k_m$ and $l_m$ to the Lagrange multipliers $\lambda_m$ and $\tau_m$ respectively.
    \item At that point, it is possible to factorize the expression over the dimension $i$ of the dataset and therefore to average over $\bx$.
    \item We expand at leading order in $N_v$ in the exponential of the results of the average taken over the dataset. For each data point $m$, it leads to a 2-variables coupled Gaussian integral over the conjugate parameters $k_m$ and $l_m$.
    \item It is now possible to integrate over the conjugate parameters $k_m$ and $l_m$.
    \item Finally, we integrate over the Lagrange parameters. The Heaviside functions are constraining the range of the integral but it is still possible to obtain an exact expression.
\end{enumerate}
The following expression is obtained
\begin{equation*}
    \Omega_P = \int_{-1}^{1} dR \exp\left(N_v\left[ 1/2 \log(1-R^2) + \alpha \log\left(1-\frac{\rm arccos R}{\pi}\right) \right]\right),
\end{equation*}
where $\alpha = P/N_v$. Again, it is possible to compute the saddle point with respect to $R$ for various values of $\alpha$. We recover that when $\alpha=0$, the entropy is dominated by $\epsilon_G=1/2$. Then, as $\alpha$ increases, the value of $\epsilon_G$ decreases until it reaches zero as $\alpha$ goes to infinity, see fig \ref{fig:R_Err} left panel.

To compute the quenched average, one can use the replica trick. The replica trick uses the following identity
\begin{equation*}
    \log(x) = \lim_{n \rightarrow 0} \frac{x^n - 1}{n}
\end{equation*}
The trick is used to compute the average of the logarithm of the entropy. In practice, we first do the computation considering that $n$ is an integer before sending it to zero, hoping that the obtained solution will be correct. Let's stress that this method is not in general a rigorous one, although it has been demonstrated in some cases to provide the correct results~\cite{talagrand2010mean,agliari2020generalized}. It is also worth mentionning that other rigorous methods to perform the quenched average exist, e.g.~\cite{agliari2020generalized}. Then, we have $n$ replicas of our perceptrons and therefore $\ws^a$,$\wt^a$ where $a=1,\dots,n$. After performing the average over the disorder, the following order parameters are introduced
\begin{align*}
    q^{ab} &= \frac{\ws^a \cdot \ws^b}{N_v}, \\
    R^{a} &= \frac{\wt^a \cdot \ws^a}{N_v},
\end{align*}
namely the overlap between the replicated students, and the overlap between the students and the teacher. To finish the computation, an ansatz should be taken for $q$ and $R$. The most simple one is the \emph{replica symmetry} ansatz where all replicas are identical. Most of the details of the computation is given in appendix \ref{sec:comp_entropy}. The overall result is given by
\begin{equation*}
    S \sim N_v {\rm max}_R \left[ \frac{1}{2}\log(1-R) \frac{R}{2} + \alpha \int \frac{dx}{\sqrt{2\pi}}e^{-x^2/2}   {\rm erfc}\left(-\sqrt{\frac{2R}{1-R}}x\right) \log \left({\rm erfc}\left(-\sqrt{\frac{2R}{1-R}}x\right) /2\right)\right];
\end{equation*}
where ${\rm erfc}$ is the complementary error function. On fig. \ref{fig:R_Err} we show the result of both the annealed and the quenched computations.
\noindent We see that in both cases we obtain the correct result when $\alpha=0$, while the annealed approximation is slightly underestimating the error for increasing values of $\alpha$. To show the correctness of the replica symmetric ansazt, it is needed to check the stability of the solution with respect to the so-called replica symmetry breaking solutions~\cite{gardner1988optimal}. This falls out of the scope of this lecture but it is important to know that in many systems, a more elaborated parametrization of the replica parameters $q^{ab}$ and $R^a$ must be found in order to obtain the correct result. 

\subsection{Capacity of the network}

Another interesting question that arises is about the capacity of the network, or to rephrase, the number of patterns that can be learned by the perceptron. In this context, we do not need a teacher, but only a set of $P$ training samples $\{\bx^{(m)},\ty^{(m)}\}$ for $m=1,\dots,P$ that will be used as patterns. We want to impose that the perceptron is capable of classifying correctly these samples (or patterns), imposing the following conditions on the student weights
\begin{equation}
    \frac{1}{\sqrt{N_v}} \ty^{(m)} \ws \cdot \bx^{(m)} \geq \kappa\;\;\; \forall m=1,\dots,P \label{eq:storageP}
\end{equation}
where we introduced $\kappa$, a positive constant ensuring a small margin of error. With this condition, we can again count the number of compatible solutions. In this case we do not have a teacher and therefore we do not need to impose any overlap with an oracle. Therefore,
\begin{equation*}
    \Omega(\{\bx^{(m)}\,\ty^{(m)}\}_m) = \int d\ws \delta(\ws^2 - N) \prod_m^P \theta\left(\frac{1}{\sqrt{N_v}} \ty^{(m)} \ws \cdot \bx^{(m)}-\kappa\right)
\end{equation*}
Using again the replica trick to compute the quenched entropy, it is possible for a given value of $\kappa$ to compute the value of the critical capacity, $\alpha_c$, below which there exists an exponential number of perceptrons capable of learning the patterns, while for higher values of $\alpha$, it is impossible. The result is given by the following condition
\begin{equation*}
    \frac{1}{\alpha_c} = \int_{-\infty}^\kappa \frac{dx}{\sqrt{2\pi}}e^{-x^2/2} (\kappa-t)^2 .
\end{equation*}
On fig. \ref{fig:R_Err} right panel, we show the behavior of the critical capacity $\alpha_c$ as a function of $\kappa$, where we can observe how it decreases as the margin of the error increases.

This concludes the main discussions on the perceptron. We saw that approaches using Statistical Physics can lead to answer interesting questions on fundamental properties of the perceptron, and a more recent analysis show that using similar tools, it is possible to analyze multilayer neural networks~\cite{10.21468/SciPostPhysCore.2.2.005}. We will now depart from supervised learning to discuss results on the RBM, a generative model.

\section{From Perceptron to Hopfield to Restricted Boltzmann Machines}
We can now make a series of connections, starting from the relation between the perceptron and the Hopfield model~\cite{hopfield1982neural}. Then, we will see how the Hopfield model can be rephrased as a Restricted Boltzmann Machine. Instead of focusing on the patterns a perceptron can learn, we can deal with another point of view. Can we define a ``simple'' dynamics over the samples/patterns $\{\bx^{(m)}\}$, such that they are both the fixed points of the dynamics and attractors? To clarify this, we can imagine that instead of having one perceptron, we could define a set of $N_v$ perceptrons, where $\bw_{i}$ are the weights of the perceptron $i$. Then using eq. \ref{eq:output_perceptron} for all the perceptrons and stacking them together, we can define the following dynamics between the output of the set of perceptrons and a new input:
\begin{equation}
    x_i(t+1) = {\rm sgn}\left(\sum_j w_{ij} x_j(t)\right). \label{eq:dyn_perceptron}
\end{equation}
Here $x_i(t)$ is the output of the perceptron $i$ at step $t$. Then, this output is re-injected as input for the next step of the dynamics at $t+1$. For each of them, we could impose that the patterns are fixed points of the dynamics, meaning that the output of the perceptron $i$ reproduces the input $i$ all the time. Hence, dropping the time indices in eq. \ref{eq:dyn_perceptron} and writing that the l.h.s. should be of the same sign as the argument of the sign function, we obtain the following set of conditions over all patterns/samples $m$:
\begin{equation*}
    x_i^{(m)} \sum_j w_{ij} x_j^{(m)} \geq \kappa \; , \;\; \forall m=1,\dots,P \text{ and } \forall i=1,\dots,N_v ,
\end{equation*}
where again, a parameter $\kappa \geq 0$ can be introduced to have a greater stability under the dynamics of eq. \ref{eq:dyn_perceptron}. We see that it is directly related to the storage problem of the perceptron by comparing it to eq. \ref{eq:storageP}. Assuming a set of orthogonal samples/patterns (in the large $N_v$ limit), $\sum_i x_i^{m} x_i^{n} \approx \delta_{nm}$, it is possible to impose them to be fixed points of eq. \ref{eq:dyn_perceptron} by setting the weight matrix as:
\begin{equation*}
    w_{ij} = \frac{1}{N_v} \sum_m x_i^{(m)} x_j^{(m)}.
\end{equation*}
This is the Hebb's rule. Plugging this weight matrix in the equations of the dynamics given by  eq. \ref{eq:dyn_perceptron}, it is easy to verify that the patterns are fixed points of the dynamics. Let's consider that the input is in the pattern $(n)$ at time $t$: $x_i(t) = x_i^{(n)}$, we can write
\begin{equation*}
    x_i(t+1) = {\rm sgn}\left(\frac{1}{N_v}\sum_{j,m} x_i^{(m)} x_j^{(m)} x_j(t)\right) \approx {\rm sgn}\left(\frac{1}{N_v}\sum_{m} x_i^{(m)} \delta_{nm} \right) \approx {\rm sgn}\left(\frac{1}{N_v} x_i^{(n)}\right) = x_i^{(n)}
\end{equation*}
Now, let's observe that the dynamical process of eq. \ref{eq:dyn_perceptron} matches the zero-temperature dynamics of an Ising model where the couplings are given by the weight matrix $w_{ij}$, defining the Hamiltonian
\begin{equation*}
    \mathcal{H} = - \sum_{i < j} w_{ij} x_i x_j 
\end{equation*}
In fact, the corresponding Ising model has been studied for a long time and is known under the name of the Hopfield model. Now, in order to make the distinction between samples and patterns, we will denote the patterns of the Hopfield model by the symbol $\xi$. In the Hopfield model, the patterns are typically drawn from the Bernoulli distribution: $\xi_i^{\mu} =  \pm 1$ with probability one-half. We will index our samples by $\mu$ (instead of $m$ to not confuse them with the magnetization), and name the dynamical variables, $\bs$, to refer to them as spins. This gives the following Hamiltonian and Boltzmann distribution
\begin{align*}
    \mathcal{H}[\bs] &= - \frac{1}{N_v} \sum_{i < j} \sum_\mu \xi_i^\mu \xi_j^\mu s_i s_j , \\
    p[\bs] &= \frac{1}{Z}\exp\left[\frac{\beta}{N_v}\sum_\mu \sum_{i<j} \xi_i^\mu \xi_j^\mu s_i s_j  \right] = \frac{1}{Z}\exp\left[\frac{\beta}{2 N_v}\sum_\mu \left(\sum_{i} \xi_i^\mu s_i \right)^2 \right].
\end{align*}
In the zero temperature limit, we have seen that the patterns are stable with respect to the dynamics of eq. \ref{eq:dyn_perceptron}. Interestingly, the model also presents non-trivial equilibrium properties for a wide range of temperatures depending on the number of patterns $P= \alpha N_v$. It is possible to understand quickly what happens in the low storage regime, when $P \ll N_v$, ($\alpha \sim 0 $). In this regime, we can approximate the partition function as follows by using the Hubbard-Stratonovitch transformation to linearize the interaction term
\begin{align}
    Z &= \int d\bmm \exp\left(- N \beta \bmm^2/2 + \sum_i \log\left[ \cosh\left( \beta\sum_{\mu} \xi_i^\mu m_\mu \right) \right] \right) \label{eq:ZHop}\\
    &\propto \exp\left(- N \beta \hat{\bmm}^2/2 + \sum_i \log\left[ \cosh\left( \beta\sum_{\mu}\xi_i^\mu \hat{m}_\mu\right) \right] \right) \text{where } \hat{m}_\mu = \frac{1}{N_v}  \sum_i \xi_i^\mu \tanh\left[ \beta\sum_{\mu} \xi_i^\mu m_\mu \right] \nonumber
\end{align}
The dominant solutions to this equation can be found by choosing the projected magnetization $\hat{\bmm}$ to be one along a given pattern and zero along the others. This gives the following mean-field self-consistent equation:
\begin{equation*}
    m = \tanh(\beta m)    
\end{equation*}
which implies a second order phase transition at $\beta_c=1$, as one recalls from the Curie-Weiss model\footnote{Other solutions can be found, for instance recalling more patterns in the same proportion~\cite{coolen2005theory}, but they are not dominating in free energy}. In this simple case, the Hopfield model behaves as a ferromagnet where the equilibrium low temperature phase corresponds to the retrieval of one pattern among all the possible ones. 

\begin{figure}
    \centering
    \includegraphics[scale=0.8]{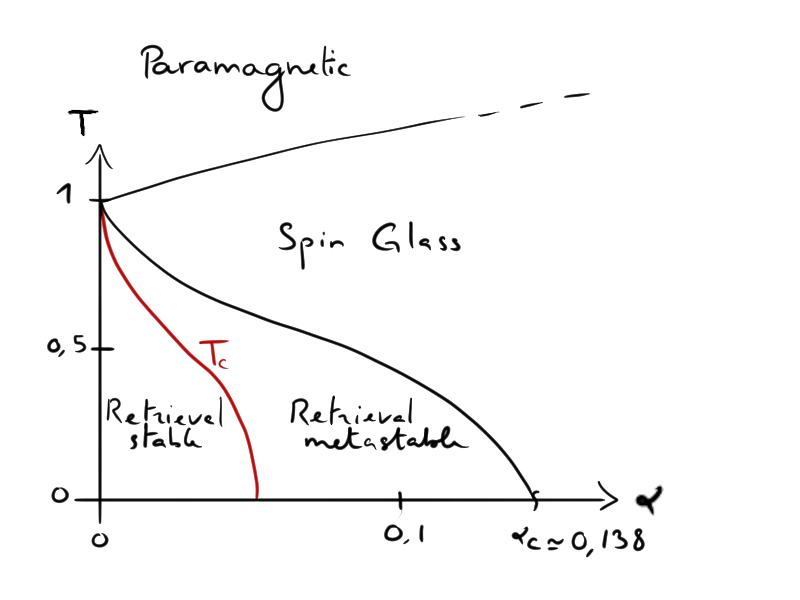}
    \caption{Phase diagram of the Hopfield model. For low values of $\alpha$, the critical temperature is equal to one, as in the Curie-Weiss model. When it starts to be proportional to the number of spins, two interesting phases arise: the retrieval where the pattern are stable and the one where they are metastable.}
    \label{fig:pd_hopfield}
\end{figure}

When the number of pattern becomes proportional to the system size $P=\alpha N_v$, we have to take care of possible correlations between the patterns. In that case indeed, a possible treatment is to use the replica trick to compute the quenched free energy. We pass the details of the computation that can be found in~\cite{nishimori1980exact,coolen2005theory} and comment quickly the phase diagram that is reproduced on fig. \ref{fig:pd_hopfield}.
We now have many interesting regions. When $\alpha$ is below a given threshold, the equilibrium configurations of the system do correspond to the patterns: it is the retrieval stable region. Then, at a given temperature, when increasing $\alpha$, the system will enter into a regime where a pattern or a mixture of patterns are metastable states of the system. In this retrieval metastable region, if the system is nearby one of these configurations, for instance close to a given pattern, it remains trapped there forever. Yet, the true equilibrium state of the system is a spin glass regime. Then for higher $\alpha$ and low-temperature, the patterns are no longer metastable and the system can only be in a spin glass phase. In a nutshell, the spin glass phase is here characterized by the presence of many metastable minima uncorrelated with the patterns. But even worst, the true equilibrium configurations are also uncorrelated to the patterns, and the system's dynamics are strongly slowed down due to the proliferation of minima. It is to be noted that, the transition in temperature at $\alpha \sim 0$ is second order, while the transition between the retrieval stable region and the spin glass one is first order.

The phase diagram of the Hopfield model is interesting as it explains how orthogonal patterns can be stored and retrieved by modelling the equilibrium states of a Boltzmann distribution. However, the fact that the interactions between the spins are pairwise, limits the ``expressivity'' of this model. In particular, if we were to use the Hopfield model as a generative model for a complex dataset, it will be impossible to adjust for the higher-moments of the data distribution. Still, by using the transformation we used to compute $Z$ in eq. \ref{eq:ZHop}, we can write the probability distribution of the Hopfield model using a Hamiltonian defined on a bipartite network,
\begin{equation}
    p[\bs,\bmm] = \frac{1}{Z} \exp\left( -N \beta \bmm^2 /2 + \beta \sum_{i,\mu} s_i \xi_i^{\mu} m_\mu\right)
\end{equation}
One layer is composed by a set of discrete variables $s_i$, and the other one by continuous Gaussian ones $m_\mu$, that we introduced earlier to linearize the Hopfield's Hamiltonian. Note in particular that there are no interactions amongst the spins or amongst the Gaussian variables. 

Starting from the same topology (a bipartite network where only half is related to the patterns we want to retrieve), we can generalize it by adjusting the type of variable that we are considering, this is what defines the RBM. The prior distribution that we choose for the $s_i$ will control the way we model our data, while the prior we choose over the variables $m_\mu$ will impact the effective Hamiltonian we obtain after integrating over them. In the classical RBM's definition, the variables $m_\mu$ are usually either discrete or distributed according to a truncated Gaussian distribution. We illustrate here why taking discrete values is an improvement over the Hopfield model. Recall that if we choose the $m_\mu$ to be Gaussian, we will recover the Hopfield model. Now, if we choose the $m_\mu$ to be discrete $\{\pm 1\}$, the effective Hamiltonian obtained after marginalizing these variables is
\begin{align*}
    \mathcal{H}_{\rm eff}[\bs] &= \sum_\mu \log\left[ \cosh( \sum_i \xi_i^{\mu} s_i) \right] \approx \sum_\mu \log\left[ 1+\frac{1}{2}\left(\sum_i \xi_i^{\mu} s_i\right)^2 + \frac{1}{4!}\left(\sum_i \xi_i^{\mu} s_i\right)^4 + \dots \right] \\
    &\approx \frac{1}{2}\sum_{i,j} s_i \sum_\mu \xi_i^\mu \xi_j^\mu s_j - \frac{7}{4!} \sum_\mu \sum_{i,j,k,l} s_i s_j s_k s_l \xi_i^{\mu} \xi_j^{\mu} \xi_k^{\mu} \xi_l^{\mu} + \dots
\end{align*}
By expanding the $\log \cosh$ in the above equation, we see that at the second order in $\bm{\xi}$ we recover the Hopfield model. But now, higher orders are present and the weights $\xi_i^{\mu}$ can be adjusted to fit best the considered dataset\footnote{Note that we do not have explicit odd-order in the expansion because of the absence of external fields in the original Hamiltonian. Odd terms would appear in a more general case}. We will now focus on the thermodynamics properties and the learning equations of such a machine.

\section{Restricted Boltzmann Machine: a generative model of spins}
\label{sec:rbm}

We understand now that the RBM can be progressively derived from the Hopfield model. Let's recall its Hamiltonian as defined in sec. \ref{ssec:rbm}
\begin{equation}
  \mathcal{H}[\bs,\btau] = -\sum_{i,\mu} s_i w_{i\mu} \tau_\mu - \sum_i s_i \theta_i - \sum_a  \tau_\mu \eta_\mu.
\end{equation}
In this interpretation, the weights of the RBM can be interpreted as patterns that are learned. In the following, we will see into more detail the learning equations of the RBM, and derive its phase diagram in a simplified mean-field context. An (oriented) recent review on the subject can be found here~\cite{decelle2021restricted}. For the rest of the section, we will note the hidden variables $\btau$ (instead of $m_\mu$), the weight matrix $w_{i \mu}$ and keep the variables to be discrete.

\subsection{Learning an RBM}

We will derive the learning equations of the RBM for discrete $\{0,1\}$ variables for both the visible and hidden layers. The generalization to other prior distributions is quite straightforward.  Let's first introduce the following notations. The average over the RBM's distribution will be noted as
\begin{equation}
    \langle f(\bm{s},\bm{\tau}) \rangle_\mathcal{H} = \sum_{\{s,\tau\}} p(\bm{s},\bm{\tau})  f(\bm{s},\bm{\tau})
\end{equation}
\noindent Before entering more into the technical details about the RBM, it is important to recall that it has been designed as a ``learnable'' generative model in practice. In that sense, the usual procedure is to feed the RBM with a dataset, tune its parameters $\bw$, $\bt$ and $\bm{\eta}$ such that the equilibrium properties of the learned RBM reproduce faithfully the correlations (or the patterns) present in the dataset. In other words, it is expected that the learned model is able to produce new data statistically similar but distinct from the training set. To do so, the classical procedure is to proceed with a stochastic gradient ascent of the likelihood function that can be easily derived. Usually, the learning of ML models involves the minimization of a loss function which happens here to be minus the log likelihood, which we will minimize by means of the classical Stochastic Gradient Descent (SGD) algorithm. First, consider a set of data points $\{s_i^{(d)}\}$, where $d=1,\dots,M$ is the index of the data. The log-likelihood is given by
\begin{align}
    \mathcal{L} & = \frac{1}{M} \sum_{d=1}^M \log\left( p(\bm{s}^{(d)} )\right) = \frac{1}{M} \sum_{d=1}^M \log\left( \sum_{\{\bm{\tau}\}} p(\bm{s}^{(d)},\bm{\tau})\right)   \nonumber \\
    & =\frac{1}{M}  \sum_{d=1}^M\left[ \log\left(\sum_{\tau=\{0,1\}} \exp\bigl(-\mathcal{H}[\bm{s}^{(d)},\bm{\tau}]\bigr)\right)\right] - \log(Z)  \nonumber \\
    & =\frac{1}{M}  \sum_{d=1}^M\left[ \sum_i \theta_i s_i^{(d)} + \sum_\mu \log\left(1 + \exp\bigl(\sum_{i} s_i^{(d)} w_{i\mu} + \eta_\mu \bigr)\right) \right] - \log(Z) \nonumber
\end{align}

\noindent Note here that in the usual definition of the RBM, external fields $\bt$ and $\bm{\eta}$ are present. The gradient w.r.t. the different parameters will then take a simple form. Let us detail the computation of the gradient w.r.t. the weight matrix. By deriving the log-likelihood w.r.t. the weight matrix we get
\begin{align}
    \frac{\partial \mathcal{L}}{\partial w_{i\mu}} &= \frac{1}{M} \sum_{d=1}^M \frac{s_i^{(d)}}{1+\exp(\sum_{i} s_i^{(d)} w_{i\mu} + \eta_\mu)} - \langle s_i \tau_\mu \rangle_\mathcal{H} \nonumber \\
    & = \frac{1}{M} \sum_{d=1}^M  s_i^{(d)} p(\tau_\mu=1|\bm{s}^{(d)}) - \langle s_i \tau_\mu \rangle_\mathcal{H} \nonumber \\
    & = \langle s_i \tau_\mu\rangle_{\rm data} -  \langle s_i \tau_\mu \rangle_\mathcal{H} \label{eq:sgd:w}
\end{align}

\noindent where we used the following notation
\begin{equation}
    \langle f(\bm{s},\bm{\tau}) \rangle_{\rm data}  = \frac{1}{M} \sum_{d=1}^M f(\bm{s}^{(d)},\bm{\tau}) p(\bm{\tau}|\bm{s}^{(d)}),
\end{equation}

\noindent that is the average over the dataset samples. The gradients for the biases (or magnetic fields) can be obtained similarly
\begin{align}
    \frac{\partial \mathcal{L}}{\partial \theta_i} &= \langle s_i\rangle_{\rm data} -  \langle s_i \rangle_\mathcal{H} \label{eq:sgd:theta} \\
    \frac{\partial \mathcal{L}}{\partial \eta_\mu} &= \langle \tau_\mu\rangle_{\rm data} -  \langle \eta_\mu \rangle_\mathcal{H} \label{eq:sgd:eta}
\end{align}

\noindent It is interesting to note that, in expression (\ref{eq:sgd:w}), the gradient is very similar to the one obtained in the traditional inverse Ising problem with the difference that in the inverse Ising  problem the first term on the r.h.s. (sometimes coined ``positive term'') depends only on the dataset, while for the RBM, we have a dependence on the model through the hidden variables (yet simple to compute), see the appendix \ref{sec:invIsing} for a quick introduction of the Ising inverse case. Once the gradient is computed, the parameters of the model are updated following the rules
\begin{align}
    w_{i\mu}^{(t+1)} &= w_{i\mu}^{(t)} + \gamma  \frac{\partial \mathcal{L}}{\partial w_{i\mu}}\Bigr|_{w_{i\mu}^{(t)},\theta_{i}^{(t)},\eta_\mu^{(t)}} \label{eq:grad:w} \\
    \theta_i^{(t+1)} &= \theta_i^{(t)}+ \gamma  \frac{\partial \mathcal{L}}{\partial \theta_{i}}\Bigr|_{w_{i\mu}^{(t)},\theta_{i}^{(t)},\eta_\mu^{(t)}} \label{eq:grad:theta} \\ 
    \eta_\mu^{(t+1)} &= \eta_\mu^{(t)}+ \gamma  \frac{\partial \mathcal{L}}{\partial \eta_{\mu}}\Bigr|_{w_{i\mu}^{(t)},\theta_{i}^{(t)},\eta_\mu^{(t)}} \label{eq:grad:eta}
\end{align}

\noindent where $\gamma$, called the learning rate, tunes the speed at which the parameters are updated in a given direction, the superscript $t$ being the  index of iteration.

The difficulty to train an RBM lies in the computation of the second term of the gradient, the so-called ``negative term'', which represents  the correlation between a visible node $i$, and a hidden node $a$, under the RBM distribution. Depending on the value of the parameters of the model (the couplings and the biases), we can either be 
\begin{enumerate}
\item in a phase where it is easy to sample configurations from  $p(\bm{s},\bm{\tau})$, --- usually called paramagnetic phase --- 
\item in a spin glass phase (if unlucky), where it is exponentially hard to escape from the spurious free energy minima; 
\item in a "recall" phase where the dominant states correspond to data-like configurations.
\end{enumerate}
But even in the latter case, it might be difficult to compute the negative term when the free energy of the model presents many ``good'' minima. In fact, to transit from one state to another one with local stochastic moves, the time is exponential in the size of the free energy barrier which is extensive in the system's size. This makes the task of sampling the model very difficult, in particular it is a phenomenon present also in simpler models, such as for instance the Hopfield model.

\subsection{Sampling and Approximation of the negative term}
Before finishing with the learning algorithm of the RBM, we need to approach into a bit more detail how the sampling of this machine is done. In practice, any Markov Chains Monte Carlo could be used to sample the equilibrium configurations of the RBM. This task is not only important to generate samples from  properly trained RBM, but it is also necessary to evaluate the ``negative term'' of the gradient. Interestingly, the architecture of the RBM gives us a very simple heat-bath algorithm to perform MC iterations. First, let's remark that the conditional probability of the visible nodes, when fixing the hidden ones (or vice-versa), factorizes as
\begin{align}
	p(\bs|\btau) = \prod_i \frac{1}{1+\exp(\sum_a w_{i\mu} \tau_\mu + \theta_i)} \text{ and therefore } p(s_i|\btau) = {\rm sig}(\sum_\mu w_{i\mu} \tau_\mu + \theta_i) \label{eq:ps_t} \\
	p(\btau|\bs) = \prod_\mu \frac{1}{1+\exp(\sum_i w_{i\mu} s_i + \eta_\mu)}  \text{ and therefore } p(\tau_\mu|\bs) = {\rm sig}(\sum_i w_{i\mu} s_i + \eta_\mu) \label{eq:pt_s} 
\end{align}
This property allows us to design a block-sampling algorithm that is commonly used for RBMs. First, you define the initial condition for the visible nodes $\bs$ (for instance, uniformly at random). Then, using eq. \ref{eq:pt_s} we can sample directly all hidden nodes independently of each other using their Bernoulli distribution. Then, the same can be done to sample back the visible nodes, fixing the value of the hidden variables to the previsouly obtained values and sampling new states for the visible ones using eq. \ref{eq:ps_t}. Since this procedure respects the detailed balance, we are (in theory) sure that we would reach equilibrium configurations by running this MC algorithm long enough.

The negative term of the gradient can therefore be computed using various MCMC chains in parallel, updating each of them with the block sampling method, which is very efficient since most of the operation can be parallelized. In practice, different schemes have been designed to reduce the number of MC steps needed for convergence. We list below the most classical ones with their eventual drawback.

\begin{itemize}
\item \textbf{Contrastive Divergence (CD)-k}: this method, introduced by Hinton~\cite{hinton2002training}, consists in initializing the MC chains on the data points of the minibatch to compute the gradient, and then to perform $k$ MC steps (with $k$ of order 1 usually). If this method does manage to learn some relevant features, we should warn that the learned RBM cannot be used at all for sampling when $k$ is too small (which is the usual setup) and therefore to generate new data uncorrelated from its initial condition. In fact, when the machine starts to learn, the number of steps is too small to depart from its initial condition and therefore the chains are very biased toward it.
\item \textbf{Persistent CD (PCD)-k}: this is an improvement over CD~\cite{tieleman2008training}. The first MC chains are started from random initial conditions. Then, the final states of the chains are reused as the initial conditions for the next computation of the gradient. So, in practice, we are keeping always the same chains (there is not reset), and at each computation of the gradient, $k$ steps are performed. For this learning dynamics, it has been observed~\cite{decelle2021equilibrium} that the learned RBM equilibrium's properties match well the empirical distribution of the dataset. New data from the learned machine will indeed be very similar to the data of the dataset. However, the quality of the generated samples will depend on the value of $k$, the larger the better, obviously.
\item \textbf{Random (Rdm)-k}: this is the most naïve way to compute the negative term, where the MC chains are started from random initial conditions at each update of the gradient, and then $k$ steps are performed. If this method is doomed to fail as soon as $k$ gets lower than the MC mixing time -- the time it takes to converge to equilibrium, --- it has some very interesting biases. In fact, when the mixing time of the system increases (which arrives eventually during the learning), the chains can not reach equilibrium configurations~\cite{decelle2021equilibrium}. At that point, an interesting phenomenon occurs, which is that the RBM learns to match the correlations of the dataset (and response of the hidden nodes) when repeating the exact same dynamics. In brief, when keeping the exact same dynamics (fixed $k$ and the distribution used for the initial conditions), the RBM will learn a dynamical process: starting from the same precise initial conditions and performing the same number $k$ of MC steps, the samples converge towards good samples. Sampling for longer time will end up in eventually reaching the equilibrium configurations of the learned model, which usually represent poorly the dataset. Of course, it is also possible to use this dynamics taking $k$ very large such that the training is equilibrated during the whole training. In that case, the model is correctly trained but it is in general very costly.

\end{itemize}

Before concluding, we should also mention an approach more familiar in Physics. The negative term can also be approximated by a mean-field estimation of the correlations. A simple mean-field approach here would be to use a high-temperature --- or small weights --- approach~\cite{plefka1982convergence,gabrie2015training,tramel2018deterministic}. In this expansion, the mean-field magnetizations of the visible and hidden nodes are given by the solutions of the self-consistent equations

\begin{align}
	m_i^{(v)} &= {\rm sig}\left( \sum_a w_{i\mu} m_\mu^{(h)} + \theta_i\right) \label{eq:mf_vis} \\
	m_\mu^{(h)} &= {\rm sig}\left( \sum_i w_{i\mu} m_i^{(v)} + \eta_a\right) \label{eq:mf_hid} 
\end{align}
Therefore, using the mean-field approximation of the free energy to compute the derivative of $\log(Z)$, we end up with 
\begin{equation}
	\langle s_i \tau_\mu \rangle_{\mathcal{H}} \approx m_i^{(v)} m_\mu^{(h)}
\end{equation}
where the mean-field values of the magnetizations are obtained by solving iteratively the eqs. \ref{eq:mf_vis}-\ref{eq:mf_hid}. This approximation can also be integrated in the different approaches developed for Monte Carlo methods: CD-MF, PCD-MF or Rdm-MF. In practice, the same limitations discussed above apply for the mean-field case. With the learning rules discussed in this section, we shall be able to learn any dataset that is binary-discrete using the algorithm \ref{Algo:RBM}.
\begin{algorithm}
    \caption{RBM learning}\label{Algo:RBM}
    \begin{algorithmic}
        \Statex \textbf{Input:} Data: $\boldsymbol{X} \in \mathbb{R}^{N\times D}$, hyperparameters: $\bm{\Theta} = N_h, \gamma, \rm{MB}$, initialization: $\bm{w}^{(t=0)} \sim \mathcal{N}(0,\sigma)$, $\bm{\eta}^{(t=0)} = \bm{0}$, $\bm{\theta}^{(t=0)} = \bm{0}$.
        \Statex \textbf{Output:} $\bm{w}^{(T)}$, $\bm{\eta}^{(T)}$, $\bm{\theta}^{(T)}$.
        
        \While{$t \le T$}
        
            \State Compute l.h.s term of the gradient, eq. \ref{eq:sgd:w}-\ref{eq:sgd:eta} using the RBM at $t$
            
            \State Compute the r.h.s. of eq. \ref{eq:sgd:w}-\ref{eq:sgd:eta} using your preferred Monte Carlo approximation iterated for $k$ steps
            
            \State Update the weights and biases using eq. \ref{eq:grad:w}-\ref{eq:grad:eta}          
            \State $t \leftarrow t+1$
        \EndWhile
    \end{algorithmic}
\end{algorithm}

This is all about the basic practical aspects of the learning part. More works on the dynamical aspects can be found here~\cite{harsh2020place,decelle2021equilibrium,decelle2018thermodynamics}, where both theoretical and numerical behaviors are inspected. In addition, a recent result about the convergence problem of the Monte Carlo chains is reported~\cite{bereux2022learning} explaining how to tackle it using a biased Monte Carlo method. Still, at a theretical level, the long term behavior of the dynamics for complex dataset is not well understood.

\subsection{Phase Diagram of RBM}

Deriving the phase diagram of RBM is not a simple question. Amongst the many reasons, a strong one is that we need to decide under what distribution of the weight matrix we want to compute the quenched free energy. A classical approach when dealing with disordered systems is to consider that the quenched variables, here the weights (or patterns in the Hopfield model), are i.i.d. from the same distribution, e.g. Gaussian variables or discrete $\pm 1$. Similar approaches have been studied for RBMs~\cite{tubiana2017emergence,barra2018phase,alemanno2022supervised} allowing the computation of a phase diagram. Yet, we can argue that during the learning, it is rather unlikely that the weights remain independent after many epochs. In these lecture notes, we will rather adapt the formalism from~\cite{decelle2018thermodynamics}, which departs a bit from the classical approach. Instead of considering independent weights, we will rather use the following assumption of a low rank-K matrix decomposition for the weight matrix $\bm{w}$ --- as if doing a singular value decomposition (SVD) and keeping the first $K$ modes ---
\begin{equation*}
    w_{i\mu} \approx \sum_{\alpha=1}^K u_i^\alpha w_\alpha v_\mu^\alpha
\end{equation*}
The idea is to consider that the eigenvalues $w_\alpha$ are encoding some intrinsic properties of the dataset while we will be averaging over the rotation matrices $\bm{u}$ and $\bm{v}$. Using this decomposition, we can rewrite the replicated partition function 
\begin{align}
    Z^n &= \sum_{\bm{s},\bm{\tau}} \exp\left[ \sum_{a=1}^n \sum_{i,\mu,\alpha} s_i^a u_i^\alpha w_\alpha v_\mu^\alpha \tau_\mu^a + \dots \right] =  \sum_{\bm{s},\bm{\tau}}  \exp\left[ \sum_{a,\alpha} w_\alpha (\sum_{i} s_i^a u_i^\alpha) (\sum_\mu  v_\mu^\alpha \tau_\mu^a) + \dots  \right] \nonumber \\
    &= \int \prod_{\alpha,a} \frac{dm_\alpha^a d\bar{m}_\alpha^a}{2 \pi} \sum_{\bm{s},\bm{\tau}} \prod_a \exp\left[ -L  \sum_\alpha \left( w_\alpha (m_\alpha^a \bar{m}_\alpha^a - m_\alpha^a s_\alpha^a  - \bar{m}_\alpha^a \tau_\alpha^a) + \eta_\alpha s_\alpha^a + \theta_\alpha \tau_\alpha^a \right) \right] \label{eq:RBM_Zp}
\end{align}
where we introduced the following quantities
\begin{align*}
    L &= \sqrt{N_v N_h}, \;\;\; \text{ and later we will also use} \;\;\; \kappa = \frac{N_h}{N_v} \\
    s_\alpha^a &= \frac{1}{\sqrt{L}} \sum_i s_i^a u_i^\alpha \;\;\; \text{ and } \;\;\; \eta_\alpha = \frac{1}{\sqrt{L}} \sum_i \eta_i u_i^\alpha \\
    \tau_\alpha^a &= \frac{1}{\sqrt{L}} \sum_\mu \tau_\mu^a v_\mu^\alpha \;\;\; \text{ and } \;\;\; \theta_\alpha = \frac{1}{\sqrt{L}} \sum_\mu \theta_\mu v_a^\alpha 
\end{align*}
We project our spin variables over the SVD decomposition of the weight matrix. Using the hypothesis that the rotation matrices $\bm{u}$ and $\bm{v}$ are made of i.i.d. elements and identifying the linear terms in $n$, the number of replicas, we get the following expression for the quenched free energy (see appendix \ref{sec:FE_RBM} for details) and the following order parameters
\begin{align*}
    f[m,\bar{m}] &= \sum_\alpha w_\alpha m_\alpha \bar{m}_\alpha - \frac{1}{\sqrt{\kappa}} \mathbb{E}_u \left[ \log 2 \cosh(h(u)) \right] - \sqrt{\kappa} \mathbb{E}_v \left[ \log 2 \cosh(\bar{h}(v)) \right] \\
    m_\alpha &= \kappa^{\frac{1}{4}} \mathbb{E}_v \left[ v^\alpha \tanh(\bar{h}(v)) \right] \;\;\; \text{ and } \;\;\; \bar{m}_\alpha = \kappa^{-\frac{1}{4}} \mathbb{E}_u \left[ u^\alpha \tanh(h(u)) \right]
\end{align*}
where we defined
\begin{equation*}
    h(u) =  \kappa^{\frac{1}{4}} \left(\sum_\gamma (w_\gamma m_\gamma-\eta_\gamma)u^\gamma \right)  \;\;\; \text{ and } \;\;\;  \bar{h}(v) =  \kappa^{-\frac{1}{4}} \left(\sum_\gamma (w_\gamma m_\gamma-\theta_\gamma)v^\gamma \right) 
\end{equation*}
Now, assuming a Gaussian distribution for the $\bm{u}$ and $\bm{v}$, we can simplify even more the expressions of the order parameters:
\begin{align*}
    m_\alpha  &= (w_\alpha \bar{m}_\alpha - \theta_\alpha)(1-q) \\
    \bar{m}_\alpha  &= (w_\alpha m_\alpha - \eta_\alpha)(1-\bar{q}) \\
    q &= \mathbb{E}_v \left[ \tanh^2(\bar{h}(v)) \right] \\
    \bar{q} &= \mathbb{E}_u \left[ \tanh^2(h(u)) \right]
\end{align*}
The interpretation of these order parameters is clear. The magnetizations $m_\alpha$ and $\bar{m}_\alpha$ represent the macroscopic magnetizations of the system projected along the $K$ modes of the weight matrix. The parameters $q$ and $\bar{q}$ represent the overlap (or spin glass order parameters) of the visible and hidden nodes respectively. From these equations, we can identify a trivial solution corresponding to the existence of a paramagnetic phase, where $m_\alpha = \bar{m}_\alpha = q = \bar{q} = 0$. An interesting question is whether this phase is stable or not in general, or if there exists a ferromagnetic phase where the magnetizations are not zero. To answer it, we can compute the Hessian of the free energy to check the stability of the solution with respect to the ferromagnetic order parameters. In such case, we obtain the following equations for the determinant of the Hessian
\begin{equation*}
    (1-q)(1-\bar{q})w_\alpha^2 = 1
\end{equation*}
Now, starting from the paramagnetic phase we fix $q = \bar{q} = 0$. We obtain that the paramagnetic phase is stable up to $w_\alpha = 1$. For higher values of $w_\alpha$, the paramagnetic phase is unstable toward the ferromagnetic order resulting in a phase transition and this determines the equivalent critical temperature of the model. Hence, under the hypothesis of low rank matrix $\bm{w}$ and the independence of the elements of the rotation matrices $\bm{u}$ and $\bm{v}$, it indicates that a transition from a paramagnetic regime to a ferromagnetic one exists. The different phases can be studied with more details, while being out of the scope of these lecture's notes, let me comment however an important result. The macroscopic behavior of the ferromagnetic phase depends on the distribution taken for the rotation matrices. In fact, when using the Gaussian distribution, it is possible to show that only the condensation towards the strongest mode --- the one with the highest $w_\alpha$ --- is stable. It is possible to obtain a mixture of modes by considering distributions with higher kurtosis. 

\begin{figure}
	\centering
    \includegraphics[scale=0.5]{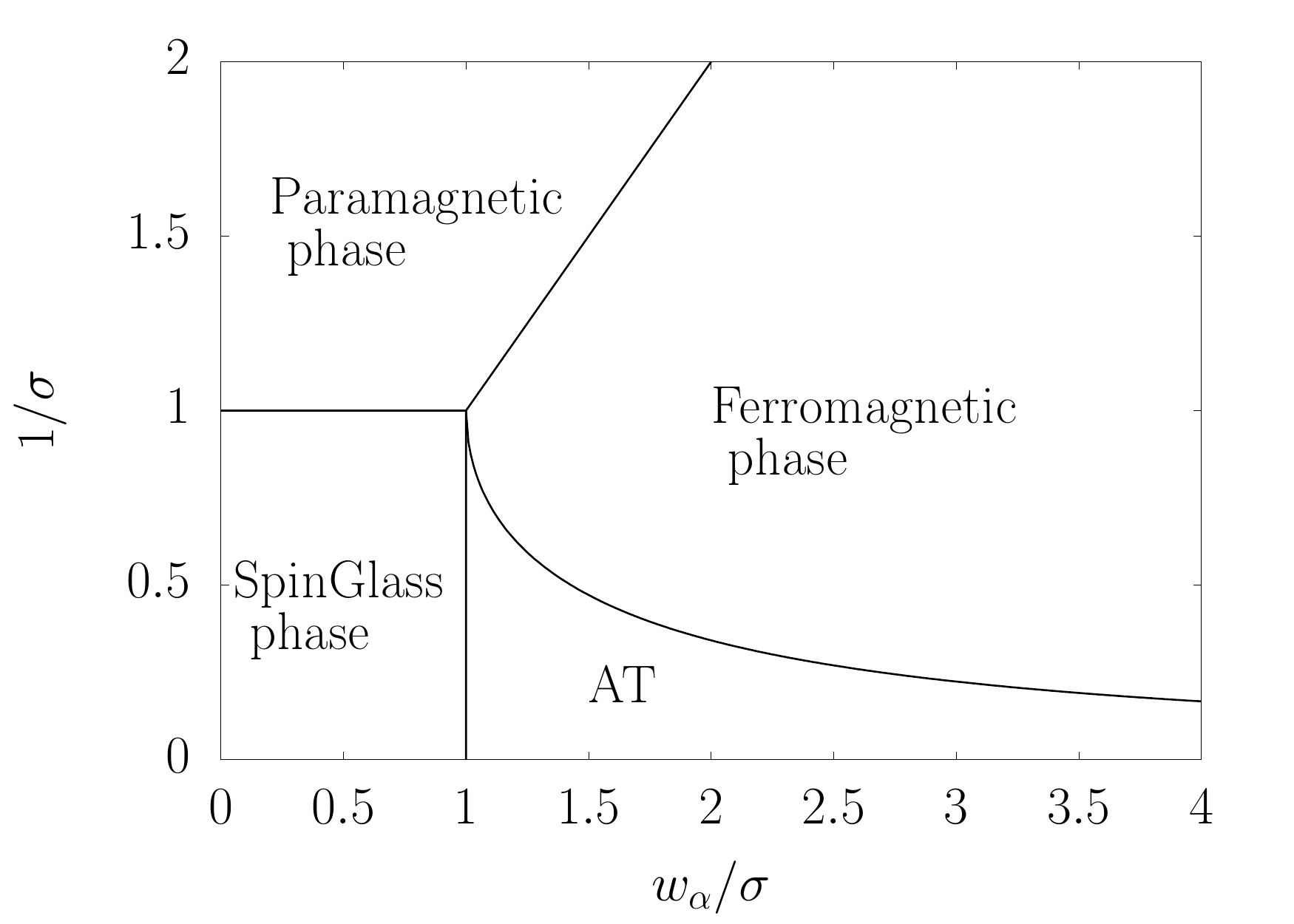}
	\caption{Phase diagram of the mean-field theory, assuming a weight matrix of the form of eq. \ref{eq:w_all}. The presence of a strong noise $\sigma$ in this model show that the system can be in a spin glass phase, while the ferromagnetic phase is dominated by the amplitude of the strongest mode $w_\alpha$ .}
	\label{fig:PT_RBM}
\end{figure}

\paragraph{Full phase diagram:} a more detailed phase diagram can be obtained by introducing noises in the weight matrix:
\begin{equation}
    w_{i\mu} \approx \sum_{\alpha=1}^K u_i^\alpha w_\alpha v_\mu^\alpha + r_{i \mu}
    \label{eq:w_all}
\end{equation}
The elements of $r_{i\mu}$ are independent Gaussian random variables with zero mean and variance $\sigma$. Following the analysis as in~\cite{decelle2018thermodynamics}, the phase diagram of fig. \ref{fig:PT_RBM} is obtained illustrating the three possible phases: paramagnetic, ferromagnetic and spin glass.
Again, without entering into details of the spin glass theory, the spin glass phase corresponds to a dynamically slow regime made of many metastable states, where the equilibrium properties are uncorrelated with the principal directions of the weight matrix and hence to the principal directions of the dataset.

This concludes the last part on RBM. For this model, physicists have brought interesting new directions to characterize the static behavior of the RBM as a function of its parameters. Yet, despite many works on this machine, most of its learning mechanisms remain poorly understood when dealing with complex (hence real) datasets, and therefore leave much space for future works. For instance, a series of open questions remain:
\begin{itemize}
  \item How the learning behavior is related to the dataset considered. We know that at the beginning of the learning, the covariance matrix is driving the learning process. It is not clear at all, for long learning time, how and when the features are shaped by the gradient.
  \item How the number of hidden nodes affect the learning process. When too few hidden nodes are present, we clearly obtain a ``limited machine'' as expected. When increasing this number, we obtain better results but we have no idea how to estimate the correct number of nodes that should be used.
  \item This relates also to the more general question of measuring the performance of generative models, yet this is not specific to the RBM.
  \item Finally, we know that the likelihood is not convex, yet we do not know if the different solutions obtained (varying the initial conditions) by the learning dynamics are similar or not, or even linked by low-free-energy paths.
\end{itemize}

\section{Conclusion}
The goal of these lecture notes was to present simple models from two aspects: Computer Science and Statistical Physics. First the perceptron, a building block of deep neural networks, which despite its simplicity can be shown to exhibit a rich behavior. It can classify, with a small error rate, non-trivial datasets and its analysis can shed light on the mechanisms taking place during the learning and on the geometrical interpretation of the learned machine. At the same time, Statistical Physics can be used to understand its space of solutions by assuming an ensemble for the dataset, opening the way to compute its capacity and generalization error for instance. Second, the Ising model, a textbook model used to study the ferromagnetism and phase transitions, is tuned into a bipartite generative model. This generative model, when learned correctly, is then capable of reproducing non-trivial correlation patterns (higher than two-body) and thus can be used to obtain very complex samples such as real images. On this model, we can observe that computer scientists have brought interesting new methods for learning efficiently the parameters, and more lately the statistical physicists have tried to characterize the phase diagram given an ensemble for the weight distribution.

At the moment where these notes are written, the field of Statistical Physics of Learning is very active, and it would be largely out of the scope to list all the relevant and interesting works that have been done recently. However, I wish to list here a non-exhaustive ``personnaly-biased'' list of works related to the study of Machine Learning by Statistical Physics approach:
\begin{itemize}
	\item Glassy behavior of deep neural networks~\cite{baity2018comparing}
	\item Learning dynamics --- gradient flow, in shallow linear neural networks~\cite{sarao2020complex}
	\item Learning behavior of neural networks on structured artificial dataset~\cite{goldt2020modeling}
	\item Why broad minima are better minima in neural networks~\cite{baldassi2016unreasonable}
	\item Computing the optimal errors in generalized linear models~\cite{barbier2019optimal}
	\item How to use more efficiently hard samples during the training~\cite{https://doi.org/10.48550/arxiv.2206.14486}
\end{itemize}

\section{Acknowledgement}
I'm indebted to the following colleagues that accepted to read this manuscript and helped me to improve it: Beatriz Seoane, Javier Moreno Gordo, Isidoro González-Adalid Pemartin. A.D. was supported by the Comunidad de Madrid and the Complutense University of Madrid (Spain) through the Atracción de Talento program (Ref. 2019-T1/TIC-13298).

\section{Appendix}
\subsection{Generalization error without training samples}
\label{sec:hypersphere}

In the computation of the generalization error before seeing training samples, we can compute the typical generalization error. The formula is given by the saddle point of the following entropy
\begin{equation*}
    \Omega_0(\epsilon) = \int d\bw_S \delta(\bw_s^2 - N_v) \delta(\frac{\bw_S \cdot \bw_T}{N_v} - R(\epsilon))
\end{equation*}
where $R(\epsilon) = \cos(\pi \epsilon)$. To compute this integral, we can use the Fourier representation of the delta distribution. We then perform a saddle point computation w.r.t. the conjugate parameter associate to the spherical constraint.
\begin{align*}
    \Omega_0(\epsilon) &= \int d\bw_S \frac{dz d\lambda}{(2\pi)^2} \exp\left[-z(\bw_S^2 - N_v) - i \lambda (\frac{\bw_S \cdot \bw_T}{N_v} - R)\right] \\
    &= \int \frac{dz d\lambda}{(2\pi)^2} \exp\left[z N_v + i \lambda R \right] \prod_i\left[ \int dw_{S,i} \exp\left(-z w_{S,i}^2 - i \lambda \frac{w_{S,i}w_{T,i}}{N_v} \right)\right] \\
    &= \int \frac{dz d\lambda}{(2\pi)^2} \exp\left[z N_v + \frac{N_v}{2}\log(\pi / z)\right] \exp\left[ -\frac{\lambda^2}{4 N_v z} + i\lambda R \right] \\
    &= \int \frac{dz}{(2\pi)^2} \sqrt{4 N_v z \pi} \exp\left[ \frac{N_v}{2}\left( \log(\pi) - \log(z) + 2z - 2R^2 z \right) \right] = \int dz g(z) \exp\left(\frac{N_v}{2} f(z) \right)
\end{align*}
where we used the fact that the teacher is a vector of the sphere $\bw_T^2 = N_v$. We can now compute the saddle point of the argument in the exponential. We find that
\begin{align*}
    z^{*} &= \frac{1}{2(1-R^2)} \\
    f(z^*) &= \log(2\pi) + \log(1-R^2) + 1
\end{align*}
which give the asymptotic behavior of the entropy. 

\subsection{Annealed and quenched Perceptron entropy}
\label{sec:comp_entropy}

\paragraph{\textbf{Annealed case ---}} 
We give some details about the computation of the annealed perceptron's entropy. We wish to compute
\begin{equation*}
    S_P^{\rm ann} = \log\left\langle \Omega\left( \bw_T,\{\bx\}_{\bw_T,\bx} \right) \right\rangle 
\end{equation*}
The main difficulty is to compute $\langle \Omega\left( \bw_T,\{\bx\}_{\bw_T,\bx} \right) \rangle= \Omega_P$. The average number of states, is given by
\begin{equation*}
    \Omega_P = \int d\bw_S \delta(\bw_S^2- N_v) \prod_m \left\langle \theta\left( \left[\frac{1}{\sqrt{N_v}}\bw_S \cdot \bx^{(m)} \right] \left[\frac{1}{\sqrt{N_v}}\bw_T \cdot \bx^{(m)} \right] \right) \right\rangle
\end{equation*}
We can use the Lagrange multipliers $\lambda_m$ and $\tau_m$ for the argument in the Heaviside function introducing delta functions and then use the Fourier representation of the $\delta$s
\begin{align*}
    \Omega_P &= \int d\bw_S \delta(\bw_S^2 - N_v) \prod_m \left\{ \frac{d\lambda_m d\tau_m dk_m dl_m}{(2\pi)^2} \theta(\lambda_m \tau_m) \left\langle \exp\left( -i k_m \left[\lambda_m - \frac{1}{\sqrt{N_v}} \bw_S \cdot \bx^{(m)} \right]  \right) \right. \right. \\
    & \left. \left. \times \exp\left[ -i l_m \left(\tau_m - \frac{1}{\sqrt{N_v}} \bw_T\cdot \bx^{(m)} \right]  \right) \right\rangle \right\}
\end{align*}
From this point, we can factorize the average over the data $p(\bx)$. In our case, independent variables uniformly distributed in $\{\pm1\}$ we can rewrite the part in the bracket as
\begin{align}
    \biggl\langle \cdot \biggl\rangle &= \exp({-ik_m \lambda_m -il_m \tau_m}) \prod_i \sum_{x_i^{(m)}} p(x_i^{(m)}) \exp\left(\frac{x_i^{(m)}}{\sqrt{N_v}} (k_m w_{S,i} +  l_m w_{T,i})\right) \nonumber \\
    &= \exp\left[-ik_m \lambda_m - il_m \tau_m + \sum_i \log \cos\left( \frac{1}{\sqrt{N_v}}(k_m w_{S,i} +  l_m w_{T,i}) \right)   \right] \label{eq:ann_cos}
\end{align}
Expanding the $\cos$ and then the $\log$ at the second order, and using the fact that $||\bw_T||^2 = ||\bw_S||^2 = 1$, we will be able to perform the integrals over the variables $k_m$ and $l_m$:
\begin{align*}
    \int \frac{dk_m dl_m}{(2\pi)^2}  \biggl\langle \cdot \biggl\rangle &= \int \frac{dk_m dl_m}{(2\pi)^2} \exp\left[ -\frac{k_m^2}{2} -\frac{l_m^2}{2} - ik_m \lambda_m - il_m \tau_m  - k_m l_m R\right] \\
    &=  \frac{1}{2\pi (1-R^2)}\exp\left( -\frac{\lambda_m^2 + \tau_m^2 - 2 \lambda_m \tau_m R}{2(1-R^2)} \right)
\end{align*}
For the last steps, we will introduce another delta function to impose that the scalar product between the student and the teacher is equal to $R$, which will decouple the integral over $\bw_S$ from the Lagrange multiplier.
\begin{align*}
    \Omega_P &= \int d\bw_S \delta(\bw_S^2 - N_v) \int_{-1}^1 dR \delta(\bw_S \cdot \bw_T - R)   \prod_m \int \frac{d\lambda_m d\tau_m}{2\pi\sqrt{1-R^2}} \theta(\lambda_m \tau_m) \exp\left(-\frac{\lambda_m^2 + \tau_m^2 - 2 \lambda_m \tau_m R}{2(1-R^2)} \right) 
\end{align*}
We focus on the integrals over the Lagrange multipliers $\lambda_m$ and $\tau_m$. The Heaviside function imposes that both parameters should be of the same sign. Now we can rewrite the whole term as
\begin{equation*}
    \frac{2}{1-R^2} \int_0^{\infty} \frac{d\lambda_m d\tau_m}{2 \pi}\exp\left(-\frac{\lambda_m^2 + \tau_m^2 - 2 \lambda_m \tau_m R}{2(1-R^2)} \right) = 1-\frac{1}{\pi} {\rm arccos} R
\end{equation*}
Putting all together and using the expression of eq. \ref{eq:perceptron_cap} we obtain in the limit of large $N_v$
\begin{equation*}
    S_P^{\rm ann} = N_v {\rm max}_R \left[ \frac{1}{2} \ln(1-R^2) + \alpha \ln\left( 1- \frac{1}{\pi} {\rm arccos} R \right) \right]
\end{equation*}
where we used the fact that $P=\alpha N_v$.

\paragraph{\textbf{Quenched case ---}} This case is much more complicated, we will however sketch a few steps to give a flavor of the type of computation and refer the reader to~\cite{engel2001statistical} for a more detailed derivation. First, we shall use the replica trick to compute this term. The replica trick is based upon the following identity
\begin{equation*}
    \langle \ln \Omega \rangle = \lim_{n \to 0} \frac{\langle \Omega^n \rangle -1}{n}
\end{equation*}
Therefore, our task is to compute the disordered average of $\Omega^n$ and then to identify the linear terms in $n$. In order to compute the average of $\Omega^n$ we will assume that $n$ is an integer:
\begin{equation*}
    \Omega^n = \int \prod_{a=1}^n d\bw_S^a \delta((\bw_S^a)^2- N_v) \prod_m \prod_{a=1}^n  \theta\left( \left[\frac{1}{\sqrt{N_v}}\bw_S^a \cdot \bx^{(m)} \right] \left[\frac{1}{\sqrt{N_v}}\bw_T \cdot \bx^{(m)} \right] \right)
\end{equation*}
We see that the same path can be taken as the annealed case, with the difference that we have now many students indexed by $a$. We therefore get, in place of eq. \ref{eq:ann_cos}, the following
\begin{align*}
    \biggl\langle \cdot \biggl\rangle =  \exp\left[ -i\sum_a k_m^a \lambda_m^a - i l_m \tau_m+\sum_i \log \cos\left( \frac{1}{\sqrt{N_v}}( \sum_a k_m^a w_{S,i}^a + l_m w_{T,i}^a) \right)\right]
\end{align*}
where we have an additional dependence in the replica of the system. The expansion of the cosine at the second order will couple the replica, and the argument in the exponential becomes
\begin{equation}
    \sum_i \log \cos\left( \frac{1}{\sqrt{N_v}}( \sum_a k_m^a w_{S,i}^a + l_m w_{T,i}^a) \right) \approx -\frac{l_m^2}{2} - \frac{1}{2}\sum_a (k_m^a)^2 - l_m \sum_a k_m^a R^a - \sum_{a<b} q^{ab} k_m^a k_m^b 
    \label{eq:replica_1}
\end{equation}
where the variables $R^a$ and $q^{ab}$ are introduced by mean of delta functions, and defined as
\begin{align*}
    R^a &= \frac{1}{N_v} \bw_S^a \cdot \bw_T \\
    q^{ab} &= \frac{1}{N_v} \bw_S^a \cdot \bw_S^b
\end{align*}
which are respectively the overlap of the student $a$ with the teacher, and the overlap between the students. Assuming that all the replicas are equivalent, and therefore $q^{ab} = R^a = R$. Rewriting eq. \ref{eq:replica_1} using this ansatz we get
\begin{equation*}
    -\frac{1}{2} (l_m \; \bm{k}_m)^T \bm{A} (l_m \; \bm{k}_m) - \bm{b}^T(l_m \; \bm{k}_m) \text{ where } A_{ab} = \delta_{ab} (1-R) + R \bm{1}_{ab} \text{ and } \bm{b} = i(\tau_m \; \bm{\lambda}_m)
\end{equation*}
where $\bm{1}$ is the matrix having all elements are equal to one, and the indices runs over $n+1$ values, since we have to include the teacher. The eigenvectors of the matrix $\bm{A}$ are $e_0 = 1+nR$ and $e_1 = 1-R$, with degeneracy $1$ and $n$ respectively. The matrix $\bm{A}$ can easily been inverted
\begin{equation*}
    \bm{A}^{-1}_{ab}  = \frac{1+nR}{(1-R)(1+nR)} \delta_{ab} - \frac{R}{(1-R)(1+nR)} \bm{1}_{ab}
\end{equation*}
Using this, we can integrate over the variables $l_m$ and $\bm{k_m}$, obtaining for a given $m$
\begin{align*}
    &\left\langle \prod_a \theta\left( \left[\frac{1}{\sqrt{N_v}}\bw_S^a \cdot \bx^{(m)} \right] \left[\frac{1}{\sqrt{N_v}}\bw_T \cdot \bx^{(m)} \right] \right) \right\rangle =\int \frac{d\tau_m}{\sqrt{2\pi (1+nR)}} \prod_a \frac{d\lambda^a_m}{\sqrt{2\pi (1-R)}}  \prod_a \theta(\lambda^a_m \tau_m) \\ & \times \exp\left[-\frac{1}{2(1-R)(1+nR)} \left( (1+(n-1)R) \tau_m^2 + (1+(n-1)R)\sum_a (\lambda_a^m)^2 - 2R\tau_m \sum_a \lambda_m^a - R\sum_{a \neq b} \lambda_m^a \lambda_m^b \right)  \right]
\end{align*}
This expression can be further simplified. First, the theta functions can be removed by changing the limits of the integrals. Then, renaming $\tau_m \rightarrow \lambda_m^{n+1}$ we can rewrite the integrals as
\begin{align*}
    \biggl\langle . \biggl\rangle &= 2\sqrt{\frac{2\pi(1-R)}{2\pi(1+nR)}}\int_0^\infty \prod_{a=1}^{n+1}\frac{d\lambda^a_m}{2\pi(1-R)} \exp\left[-\frac{1}{2(1-R)(1+nR)} \left((1+nR)\sum_a (\lambda_a^m)^2 - R(\sum_a \lambda_m^a)^2 \right) \right] \\
    &= 2\sqrt{\frac{2\pi(1-R)}{2\pi(1+nR)}}\int_0^\infty \prod_{a=1}^{n+1}\frac{d\lambda^a_m}{2\pi(1-R)} \int \frac{dt}{\sqrt{2\pi}} e^{-t^2/2} \exp\left[t\sum_a \lambda_m^a \sqrt{\frac{R}{(1-R)(1+nR)}}\right] \\
    & \times \exp\left[ -\frac{1}{2(1-R)(1+nR)} \left((1+nR)\sum_a (\lambda_a^m)^2 \right)\right]
\end{align*}
where we linearized the square within the exponential using the Hubbard-Stratonovitch transformation. We can simplify this by changing the limits of the integrals over $\lambda_m^a$. Reinserting also the product over the training samples, we get
\begin{equation*}
    \prod_m \left\langle \prod_a \theta\left( \left[\frac{1}{\sqrt{N_v}}\bw_S^a \cdot \bx^{(m)} \right] \left[\frac{1}{\sqrt{N_v}}\bw_T \cdot \bx^{(m)} \right] \right) \right\rangle = \left[ 2 \int \frac{1}{\sqrt{2\pi}} e^{-t^2/2} \left\{ \int_{-\sqrt{\frac{R}{1-R}}t}^{\infty} \frac{dx}{\sqrt{2\pi}} e^{-x^2/2} \right\}^{n+1} \right]^{\alpha N_v}
\end{equation*}
The entropic term can now be calculated. In addition to the delta function over the spherical constraint, we have also the constraints on the overlaps between the various replicas, and between the replicas and the teacher. Recall that we renamed our teacher as a $n+1$ replica of the system. Under the replica symmetric hypothesis, it simplifies the expression as follows
\begin{align*}
    S &= \int \prod_a d\bw^a \delta((\bw^a)^2 - N_v) \prod_{a<b} \delta(\bw^a \cdot \bw^b  - N_v R) \\
    &= \int \prod_a d\bw^a dz_a \prod_{a<b} dq_{ab} \exp\left[ -\sum_a z_a((\bw^a)^2 - N_v) - \sum_{a<b}(\bw^a \cdot \bw^b - N_v R) \right] \\
    &= \int \prod_a d\bw^a dz_a \prod_{a<b} dq_{ab} e^{\sum_a  z_a N_v + \sum_{a<b} q_{ab} R N_v } \prod_i \left\{\int d\bw \exp\left[ -\sum_a z_a (w^a)^2 - \sum_{a<b} q_{ab} w^a w^b \right]  \right\}
\end{align*}
We can define the matrix 
\begin{equation*}
    \bm{A}_{ab} = 2 z_a\delta_{ab} + (\bm{1}_{ab} - \delta_{ab})q_{ab}
\end{equation*}
and compute formally the Gaussian integral over the parameters $\bw$, obtaining
\begin{equation*}
    S = \int \prod_a dz_a \prod_{a<b} dq_{ab}  \exp\left[ \frac{N_v}{2} \log (2 \pi) -\frac{N_v}{2} {\rm Tr} {\rm Log} \bm{A} + \sum_a  z_a N_v + \sum_{a<b} q_{ab} R N_v \right]
\end{equation*}
We can evaluate now the whole integral using the saddle point method over the variables $z_a$ and $q_{ab}$. We get
\begin{align*}
    \frac{\partial }{\partial z_a} &= -\frac{N_v}{2} \frac{\partial A_{aa}}{\partial z_a} {A}^{-1}_{aa} + N_v  = -N_v {A}^{-1}_{aa} + N_v\\
    \frac{\partial }{\partial  q_{ab}} &= - \frac{N_v}{2} \frac{\partial A_{ab}}{\partial q_{ab}} {A}^{-1}_{ba} -\frac{N_v}{2} \frac{\partial A_{ba}}{\partial q_{ab}} {A}^{-1}_{ab}  + R N_v = -N_v A^{-1}_{ab} + R N_v
\end{align*}
From this we can compute the inverse matrix $\bm{A}^{-1}$ and its eigenvalues. The result is given by
\begin{equation*}
    S \sim \exp\left[ \frac{N_v}{2} \left( n\log (1-R) + \frac{1}{2} \log(1+nR) \right) \right]
\end{equation*}

Putting everything together, we obtain
\begin{equation*}
    \langle \Omega^n \rangle = \int dR \exp\left( N_v \left[ \frac{n}{2} \log (1-R) + \frac{1}{2} \log(1+nR) + \alpha \log\left( 2 \int \frac{1}{\sqrt{2 \pi}} e^{-t^2/2} \left\{ \int_{-\sqrt{\frac{R}{1-R}}t}^{\infty} \frac{dx}{\sqrt{2\pi}} e^{-x^2/2} \right\}^{n+1}  \right)  \right] \right)
\end{equation*}
Now, isolating the term proportional to $n$, we obtain that 
\begin{equation*}
    \frac{\langle \ln \Omega \rangle}{N_v} = {\rm max}_R \left[ \frac{1}{2}\log(1-R) +\frac{R}{2} + 2\alpha \int \frac{1}{\sqrt{2 \pi}} e^{-t^2/2} \left\{ \int_{-\sqrt{\frac{R}{1-R}}t}^{\infty} \frac{dx}{\sqrt{2\pi}} e^{-x^2/2} \right\} \log \left\{ \int_{-\sqrt{\frac{R}{1-R}}t}^{\infty} \frac{dx}{\sqrt{2\pi}} e^{-x^2/2} \right\} \right]
\end{equation*}

\subsection{Equivalence between the RBM and the Hopfield model}
\label{sec:equivRBM_HOPF}
We show the equivalence between the RBM and the Hopfield model~\cite{barra2012equivalence,mezard2017mean,marullo2020boltzmann} when having discrete binary variables for the visible nodes and Gaussian ones for the hidden layer. Let's consider the Hopfield model with $P = \alpha N_v$ binary patterns
\begin{equation*}
    p[\bs] = \frac{1}{Z} \exp\left[ \frac{\beta}{N_v} \sum_\mu \sum_{i<j} \xi_i^{\mu} \xi_j^{\mu} s_i s_j \right] = \frac{1}{Z} \exp\left[ \frac{\beta}{2N_v} \sum_\mu \left(\sum_i \xi_i^{\mu} s_i \right)^2 + {\rm cte} \right]
\end{equation*}
Using the Hubbard-Stratonovitch transformation, we can linearize the square and rewrite it as
\begin{equation*}
    p[\bs] = \frac{1}{Z} \int d\bmm \exp\left[ -\frac{\sqrt{N_v}(\bmm)^2}{2} + \sqrt{\beta} \sum_{i \mu} s_i \xi_i^\mu m_\mu\right]
\end{equation*}
Now, we can interpret the integrand as a new probability distribution over the variables $\bs$ and $\bmm$
\begin{equation*}
    p[\bs,\bmm] = \frac{1}{Z} \exp\left[ -\frac{(\bmm)^2}{2} + \sqrt{\frac{\beta}{N_v}} \sum_{i \mu} s_i \xi_i^\mu m_\mu\right]
\end{equation*}
Hence, the Hamiltonian of the RBM is related to the one of Hopfield according to the following change
\begin{equation*}
     \sqrt{\frac{\beta}{N_v}}\xi_i^\mu \rightarrow  w_{i\mu}
\end{equation*}

\subsection{Inverse Ising problem}
\label{sec:invIsing}
The inverse Ising problem consists in matching the coupling constants of an Ising model such that it fits best a given dataset. Let's consider a dataset $\{s_i^{(m)}\}$ where $m=1\dots,M$ of discrete $\pm 1$ variables, and the following Ising model
\begin{equation*}
    p[\bs] = \frac{1}{Z} \exp\left[ \sum_{i<j} J_{ij}s_i s_j + \sum_i h_i s_i\right]
\end{equation*}
We wish to adjust the variables $\bm{J}$ and $\bm{h}$ to the dataset, hence to maximize the following likelihood
\begin{equation*}
    \mathcal{L} = \frac{1}{M} \left( \sum_m \sum_{i<j} J_{ij} s_i^{(m)} s_j^{(m)} \sum_i h_i s_i^{(m)} - \log Z\right)
\end{equation*}
To maximize it, we can follow the gradient
\begin{align*}
    \frac{\partial \mathcal{L}}{\partial J_{ij}}&= \frac{1}{M} \sum_m s_i^{(m)} s_j^{(m)} - \langle s_i s_j \rangle_{\mathcal{H}} \\
    \frac{\partial \mathcal{L}}{\partial h_{i}}&= \frac{1}{M} \sum_m s_i^{(m)} - \langle s_i \rangle_{\mathcal{H}}
\end{align*}
The first term of each gradient corresponds respectively to the pairwise correlations and the magnetization given by the dataset. In particular, these terms are constant all along the learning at the opposite of the RBM where it involves the response function of the hidden nodes, involving therefore the parameters of the model. The following learning dynamics can be used
\begin{align*}
    J_{ij}^{(t+1)} &= J_{ij}^{(t)} + \gamma  \frac{\partial \mathcal{L}}{\partial J_{ij}}\Bigr|_{J_{ij}^{(t)},h_{i}^{(t)}} \\
    h_i^{(t+1)} &= h_i^{(t)}+ \gamma  \frac{\partial \mathcal{L}}{\partial h_{i}}\Bigr|_{J_{ij}^{(t)},h_{i}^{(t)}} 
\end{align*}
and the likelihood is convex in the parameters.

\subsection{Free energy of the RBM}
\label{sec:FE_RBM}
We derive the equations to obtain the mean-field free energy of the RBM. For simplicity, we do not consider the biases since they are very simple to deal with. We start by considering eq. \ref{eq:RBM_Zp}
\begin{equation*}
    Z^n= \int \prod_{\alpha,a}\frac{dm_\alpha d\bar{m}_\alpha}{2\pi} \prod_a \exp\left[ -L \sum_\alpha w_\alpha \left(m_\alpha^a \bar{m}_\alpha^a - m_\alpha^a s_\alpha^a - \bar{m}_\alpha^a \tau_\alpha^a \right) \right]
\end{equation*}
We can isolate the terms involving only $m_\alpha^a$ on one side and $\bar{m}_\alpha^a$ on the other. We obtain
\begin{equation*}
    \phi_E=\sum_{\bs,\btau}\exp\left[ L \sum_{\alpha,a} w_\alpha \left(m_\alpha^a s_\alpha^a + \bar{m}_\alpha^a \tau_\alpha^a \right)\right] = \prod_i \sum_{\{s_i^a\}}\exp\left[ \sqrt{L} s_i^a \sum_{\alpha,a} w_\alpha m_\alpha^a u_i^\alpha \right] \prod_\mu \sum_{\{\tau_\mu^a\}}\exp\left[ \sqrt{L} \tau_\mu^a \sum_{\alpha,a} w_\alpha \bar{m}_\alpha^a v_\mu^\alpha \right]
\end{equation*}
If we consider our hypothesis of i.i.d. variables for the quenched disorder $\bm{u}$ and $\bm{v}$, we obtain
\begin{align*}
    \mathbb{E}_{\bm{u},\bm{v}}& \left[\phi_E \right] = \int d\bm{u}d\bm{v} \prod_i p(u_i^\alpha) \sum_{\{s_i^a\}}\exp\left[ \sqrt{L} s_i^a \sum_{\alpha,a} w_\alpha m_\alpha^a u_i^\alpha \right] \prod_\mu p(v_\mu^\alpha)\sum_{\{\tau_\mu^a\}}\exp\left[ \sqrt{L} \tau_\mu^a \sum_{\alpha,a} w_\alpha \bar{m}_\alpha^a v_\mu^\alpha \right] \\
    &= \prod_i  \left\{ \int \prod_\alpha du_i^\alpha p(u_i^\alpha) \sum_{\{s_i^a\}}\exp\left[ \sqrt{L} s_i^a \sum_{\alpha,a} w_\alpha m_\alpha^a u_i^\alpha \right] \right\} \times \\
    & \;\;\;\;\;\; \prod_\mu \left\{ \int \prod_\alpha  dv_\mu^\alpha p(v_\mu^\alpha) \sum_{\{\tau_\mu^a\}}\exp\left[ \sqrt{L} \tau_\mu^a \sum_{\alpha,a} w_\alpha \bar{m}_\alpha^a v_\mu^\alpha \right] \right\} \\
    &= \left( \int d\bm{u} p(\bm{u}) \prod_a\sum_{\{s^a\}}\exp\left[ \sqrt{L} s^a \sum_{\alpha} w_\alpha m_\alpha^a u^\alpha \right]  \right) \left( \int d\bm{v} p(\bm{v}) \prod_a\sum_{\{\tau^a\}}\exp\left[ \sqrt{L} \tau^a \sum_{\alpha} w_\alpha \bar{m}_\alpha^a v^\alpha \right] \right) \\
    &= \mathbb{E}_{\bm{u}} \left[ 2^n \prod_p \cosh\left(\sqrt{L}\sum_{\alpha}w_\alpha m_\alpha^p u^\alpha\right) \right]^{N_v}  \mathbb{E}_{\bm{v}} \left[ 2^n \prod_p \cosh \left(\sqrt{L}\sum_{\alpha} w_\alpha \bar{m}_\alpha^p v^\alpha \right)\right]^{N_h}
\end{align*}
We can now explicitly write the scaling of the elements $u_i^\alpha \sim \tilde{u}_i^\alpha \sqrt{N_v}$ and $v_\mu^\alpha \sim \tilde{v}_\mu^\alpha \sqrt{N_h}$. By changing the integral, we obtain

\begin{align*}
    \mathbb{E}_{\bm{u},\bm{v}}\left[ Z^n \right] = \int &\prod_{\alpha,a}\frac{dm_\alpha d\bar{m}_\alpha}{2\pi} \exp\left[ -L \left( \sum_{\alpha,a} w_\alpha m_\alpha^a \bar{m}_\alpha^a -\frac{1}{\sqrt{\kappa}} \mathbb{E}_{\bm{u}}\left[\log 2 \cosh\left(\kappa^{1/4} \sum_\gamma w_\gamma m_\gamma^a u^\gamma\right)\right] \right. \right. \\
    & \left. \left. -\sqrt{\kappa} \mathbb{E}_{\bm{v}}\left[\log 2 \cosh\left(\kappa^{-1/4} \sum_\gamma w_\gamma \bar{m}_\gamma^a v^\gamma\right)\right] \right) \right]
\end{align*}
from which we can obtain the free energy by assuming the replica symmetry and gather the terms that are linear in $n$.

\bibliographystyle{unsrt}
\bibliography{biblio}

\end{document}